\documentclass[12pt,dvips]{article}
\usepackage{rotating}
\usepackage{amsmath}
\usepackage{amsthm}
\usepackage{amsfonts}
\usepackage[dvips]{epsfig}
\usepackage{graphicx}
\usepackage{amssymb}
\usepackage{cancel}
\usepackage{pstricks} 
\usepackage{lscape}
\usepackage{color}
\usepackage{cite}
\newcommand{\beq}{\begin{equation}}
\newcommand{\eeq}{\end{equation}}
\newcommand{\ga}{\lower.7ex\hbox{$\;\stackrel{\textstyle>}{\sim}\;$}}
\newcommand{\la}{\lower.7ex\hbox{$\;\stackrel{\textstyle<}{\sim}\;$}}

\begin{document}

\def\thefootnote{\fnsymbol{footnote}}

\begin{flushright}
{\tt KCL-PH-TH/2014-38}, {\tt LCTS/2014-37}, {\tt CERN-PH-TH/2014-183}  \\
{\tt ACT-11-14, MIFPA-14-28, UMN-TH-3402/14, FTPI-MINN-14/29} \\
\end{flushright}

\begin{center}
{\bf {\Large Two-Field Analysis of No-Scale Supergravity Inflation}}
\vspace {0.1in}
\end{center}

\vspace{0.05in}

\begin{center}{\large
{\bf John~Ellis}$^{a}$,
{\bf Marcos~A.~G.~Garc\'ia}$^{b}$,
{\bf Dimitri~V.~Nanopoulos}$^{c}$ \\ and \\
\vspace{0.1in}
{\bf Keith~A.~Olive}$^{b}$
}
\end{center}

\begin{center}
{\em $^a$Theoretical Particle Physics and Cosmology Group, Department of
  Physics, King's~College~London, London WC2R 2LS, United Kingdom;\\
Theory Division, CERN, CH-1211 Geneva 23,
  Switzerland}\\[0.2cm]
  {\em $^b$William I. Fine Theoretical Physics Institute, School of Physics and Astronomy,\\
University of Minnesota, Minneapolis, MN 55455, USA}\\[0.2cm]
{\em $^c$George P. and Cynthia W. Mitchell Institute for Fundamental Physics and Astronomy,
Texas A\&M University, College Station, TX 77843, USA;\\
Astroparticle Physics Group, Houston Advanced Research Center (HARC), \\ Mitchell Campus, Woodlands, TX 77381, USA;\\
Academy of Athens, Division of Natural Sciences,
Athens 10679, Greece}\\
\end{center}

\bigskip

\centerline{\bf ABSTRACT}

\noindent  
{\small Since the building-blocks of supersymmetric models include chiral superfields
containing pairs of effective scalar fields, a two-field approach is particularly
appropriate for models of inflation based on supergravity. In this paper, we generalize the
two-field analysis of the inflationary power spectrum to supergravity models 
with arbitrary K\"ahler potential. We show how two-field effects in the context of no-scale supergravity
can alter the model predictions for the scalar spectral index $n_s$ and the tensor-to-scalar ratio $r$,
yielding results that interpolate between the Planck-friendly Starobinsky model and BICEP2-friendly predictions.
In particular, we show that two-field effects in a chaotic no-scale inflation model
with a quadratic potential are capable of reducing $r$ to very small values $\ll 0.1$.
We also calculate the non-Gaussianity measure $f_{\rm NL}$,
finding that is well below the current experimental sensitivity.}

\vspace{0.2in}

\begin{flushleft}
September 2014
\end{flushleft}
\medskip
\noindent

\newpage

\section{Introduction}

The recent results from the Planck satellite~\cite{Planck} and the BICEP2 experiment~\cite{BICEP2}
have inaugurated a new era in the confrontation of inflationary models
with observation. Planck has provided whole-sky measurements of the
cosmic microwave background (CMB) with unparalleled
precision down to small angular scales, confirming the previous indications
from WMAP~\cite{WMAP} that the scalar index $n_s < 1$ and refining previous upper limits on the
tensor-to-scalar ratio $r$. Moreover, Planck has provided stringent upper limits
on many possible non-Gaussian features in the CMB~\cite{Planck}. More recently, BICEP2
has detected B-mode polarization in the CMB~\cite{BICEP2}, and triggered a raging debate on its
interpretation~\cite{debate,fix}. In particular, how much of it is due to foreground galactic dust, 
and how much may be primordial? A recent Planck analysis~\cite{Planckdust} indicates
that the BICEP2 signal may well be dominated by foreground dust,
in which case values of $r \la 0.1$ would be favoured~\cite{CHW}.

The Planck results~\cite{Planck} dampened interest in multi-field models that could give
large values of $f_{\rm NL}$ and stimulated interest in models that predicted
small values of $r$, such as the Starobinsky model based on a $R + R^2$
modification of Einstein gravity \cite{Staro,MC,Staro2}. There has also been renewed
interest in the construction of models that reduce, at least approximately,
to the Starobinsky model, such as Higgs inflation \cite{HI}. In this connection,
attention was drawn to a class of supergravity models that reduce to the
$R + R^2$ model and reproduce its predictions for $r$ and $n_s$, and
there has subsequently been much interest in exploring this class of models
\cite{ENO6,ENO7,KLno-scale,WB,FKR,ENO8,fklp,AHM,pallis,adfs,ky,klt}.

An important aspect of these models is that, since generic supersymmetric
models use chiral supermultiplets as building-blocks, and since each chiral
supermultiplet contains two real fields, generic supersymmetric models of
inflation, including supergravity models, contain an even number of scalar fields~\footnote{Note,
however, that this observation can be evaded in gauged supergravity models.}. 
For this reason, it is appropriate to use a multi-field approach to analyze the 
predictions of supersymmetric models of inflation. This has been done, for example,
in the context of a Wess-Zumino model of inflation~\cite{EMM}, and its importance for
supergravity models of inflation has been stressed in~\cite{KT}, as
was taken into account in the published version of~\cite{EGNO2}.

In supergravity models, and in two-field models in general, the evolution of the inflaton
may take a non-trivial path in field space depending on how the two fields are coupled
both in the potential and in their kinetic terms. The simplest example of such a non-trivial
path occurs when the two fields have different masses. The cosmological evolution starts down
the path of steepest decent (in the direction of the heavier field), and then turns down the
lighter direction towards the global minimum.  This case has been well studied \cite{Tsujikawa:2003,Lalak:2007vi,br},
and it has been shown that isocurvature fluctuations perpendicular to the direction of the 
motion of the scalar field eventually source adiabatic perturbations as the field evolves
towards the global minimum. The extra source of adiabatic scalar perturbations then tends to suppress
the tensor-to-scalar ratio, $r$. 

We consider the natural framework for formulating
models of inflation to be supersymmetry~\cite{ENOT,nost,hrr,gl1}, specifically local supersymmetry, i.e.,
supergravity~\cite{SUGRA}.
Whereas the Planck limits on $r$ favour inflationary models resembling the Starobinsky model~\cite{Staro,MC,Staro2},
the BICEP2 results point back to simple models of quadratic chaotic inflation \cite{m2}.
The first attempt at a chaotic inflation model in the context of supergravity was made in \cite{gl1}.
In view of the eta problem in supergravity~\cite{eta}, many of
the simplest models employ a shift symmetry in the K\"ahler potential \cite{kyy,Yamaguchi:2000vm}.
This may be realized in the context where the sneutrino plays the role of the inflaton \cite{snu},
or in relation to moduli stabilization \cite{mod}. However, no-scale supergravity \cite{no-scale} offers an alternative
solution  to address the eta problem, and has many other attractive features,
including its appearance in generic string compactifications~\cite{Witten}.  Here we
are interested in inflationary models \cite{gl2,EENOS,bg,Ant,ENO6,ENO7,ENO8,EGNO,EGNO2} based on
no-scale supergravity . 

In view of the current uncertainty in the value or $r$, which may take any
value between 0 and 0.1 or more, inflationary model-builders are enjoying
a field-day until the uncertainty is reduced. One possibility is to commit strongly
to some particular model of inflation and its predictions. Another approach is to
formulate more general frameworks that can accommodate a wider range of
predictions compatible with the present observational range of $r$. The latter
was the point of view taken in~\cite{EGNO2}, where it was shown how a class
of simple no-scale supergravity models could interpolate between the prediction
$r \sim 0.15$ of chaotic inflation in a quadratic potential and the prediction $r \sim 0.003$
of the Starobinsky model~\footnote{Another example of this approach is provided by the search for attractor solutions
that relate parametrically the two solutions \cite{KLno-scale,att}.}.

The value of $r$ in the model in~\cite{EGNO2} was varied by adjusting the {\it angle in the complex
plane} of a valley in the effective potential for the complex modulus field $T$ of
the simplest no-scale supergravity model that
provides the inflaton. The complications of two-field inflation~\cite{KT} were avoided in
this model by requiring the walls of the potential valley to be very steep,
effectively constraining the inflaton trajectory to the analogue of a narrow
``bobsleigh track".

In this paper we take a different approach. Working in the same no-scale supergravity framework,
we vary the parameter controlling the {\it width} of the potential valley. In this way,
we are able to dial up the importance of the two-field effects. As we show, they
have relatively little effect on the value of $n_s$ in this model, which is almost
always compatible with the experimental value $n_s \simeq 0.960 \pm 0.008$.
On the other hand, two-field effects may enhance the magnitude of the scalar
perturbations above the value expected naively from a single-field analysis,
thereby suppressing the tensor-to-scalar ratio $r$. This provides an independent
way to interpolate between the limits of chaotic quadratic inflation and
Starobinsky-like models that, as we verify, does not lead to large non-Gaussianities.

The layout of this paper is as follows. Section 2 first reviews the class of
no-scale supergravity models we consider, and then presents the (modest
extension of) the standard two-field formalism required to analyze this class of
models. Section~3 then applies this formalism to the no-scale supergravity
model of interest, focusing on initial conditions with ${\cal R}e~T = 0$, which
would seem in a single-field treatment to yield large values of $r$. We show
explicitly that, as the width of the potential valley increases, the trajectory of
the inflaton veers into the real direction: ${\cal R}e~T \gg {\cal I}m~T$ and $r$
assumes values closer to the Starobinsky value. Section~4 describes our
calculation of the non-Gaussianity parameter $f_{\rm NL}$, which we find to be
always small: $f_{\rm NL} \la 0.03$. Finally, Section~5 summarizes our
conclusions and prospects.

\section{Model and Formalism}

\subsection{Specification of the Model}

As discussed in~\cite{EGNO2}, the K\"ahler potential of the minimal no-scale supergravity model 
has the form~\cite{no-scale}
\beq
K\; = \; -3\ln(T+\bar{T})+\dots \, ,
\label{minimal}
\eeq
where the dots represent corrections to the K\"ahler potential due to perturbative
or non-perturbative effects, additional matter fields $\phi$, etc.~\footnote{We 
will work in Planck units $M_P^2 = 8\pi G_N=1$.}.
As already mentioned, one of the reasons for our interest in no-scale supergravity is that it is the
natural framework for the low-energy effective field theory in a generic string compactification~\cite{Witten},
where $T$ is identified as a complex modulus field. In a no-scale supergravity model
with an underlying SU$(2,1)$/SU$(2)$ $\times$ U(1) symmetry, the K\"alhler potential 
could be written as 
\beq
K \; = \; -3\ln\left(T+\bar{T} - \frac{\phi \phi^*}{3}\right) \, ,
\label{notourK}
\eeq
If we identify $T$ as the chiral (two-component) inflaton superfield,
and we assume  the following superpotential \cite{Cecotti} for our no-scale supergravity model of inflation:
\beq
W \; = \; \sqrt{\frac{3}{4}}\,\frac{m}{a}\phi(T-a) \, .
\label{ourW}
\eeq
Starobinsky-type inflation would occur if initial conditions placed $T$ along the real axis \cite{KLno-scale,ENO7}.
Although the potential is quadratic along the imaginary $T$ axis \cite{FeKR},
the evolution of the field moves it to the real axis, leading again to Starobinsky-like inflation
unless $T$ is stabilized along its imaginary direction using a higher order term such as $(T+T^*)^n$ in
the K\"ahler potential \cite{oops,EGNO,FeKR}. 
In addition this type of model is not easily generalized to allow inflation ``off-axis".

In typical orbifold string compactifications with three
moduli that are fixed by some unspecified mechanism at a high scale to be proportional,
the K\"ahler potential may be written in the following form~\cite{casas}:
\beq
K \; = \; -3\ln(T+\bar{T})+\frac{|\phi|^2}{(T+\bar{T})^3} \, ,
\label{ourK}
\eeq
up to irrelevant constants, where we consider a single matter field $\phi$ with modular weight $3$.
Here, as in \cite{EGNO2}, we use the superpotential given in Eq. (\ref{ourW}).
As discussed in~\cite{EGNO2}, the matter field $\phi$ is constrained by the exponential factor $e^K$:
\beq
V\; \propto \; e^{|\phi|^2/(T+\bar{T})^3} \; \simeq \; e^{(2a)^{-3}|\phi|^2} \, .
\label{fixphi}
\eeq
We assume that $a=1/2$, in which case $\phi \to 0$ rapidly at
the start of inflation, and the scalar potential takes the simple form
\beq
V = \frac{3 m^2}{4 a^2} |T-a|^2 \, .
\eeq
We write $T$ in terms of two real fields
$\rho$ and $\alpha$ that parameterize its real and imaginary components, respectively:
\beq
T \; = \; a\left(e^{-\sqrt{\frac{2}{3}}\rho}+i\sqrt{\frac{2}{3}}\,\alpha\right) \, ,
\label{rhosigma}
\eeq
with the following effective Lagrangian:
\beq
\mathcal{L} \; = \; \frac{1}{2}\partial_{\mu}\rho\partial^{\mu}\rho+\frac{1}{2}e^{2\sqrt{\frac{2}{3}}\rho}\partial_{\mu}\alpha\partial^{\mu}\alpha - 
\frac{3}{4}m^2\left(1-e^{-\sqrt{\frac{2}{3}}\rho} \right)^2 - \frac{1}{2}m^2\alpha^2 \, .
\label{effL}
\eeq

Moreover, the coupling between $\rho$ and $\alpha$ through the kinetic term in (\ref{effL})
yields a coupling between the curvature and isocurvature perturbations, 
resulting in an enhancement of the curvature modes at super-horizon scales
whose calculation requires a two-field analysis. 

\subsection{Two-Field Formalism}

Multi-field inflation has been discussed extensively in the literature (see, for example,
\cite{2field_cano,Tsujikawa:2003} for multi-field inflation with canonical kinetic terms, 
\cite{2field_non_cano,mukhanov:1998,Lalak:2007vi} for examples with particular non-canonical kinetic terms, 
and \cite{2field_gen} for a discussion of more general kinetic terms).
Here we present an additional review where we emphasize aspects of the generic multi-field formalism
relevant to inflationary models arising in $\mathcal{N}=1$ supergravity,
presenting the equations in a form that allows their immediate application.

Disregarding gauge interactions, the scalar $\mathcal{N}=1$ supergravity action for the scalar sector can be written as 
\beq\label{comp_act}
S \; = \; \int d^4x\, \sqrt{-g}\left[K_{I\bar{J}}\partial_{\mu}\Phi^I\partial^{\mu}\bar{\Phi}^{\bar{J}} - V(\boldsymbol{\Phi}) \right] \, ,
\eeq
with the scalar potential
\beq\label{sugra_pot}
V \; = \; e^{K}(K^{\bar{I}J} D_{J}W\bar{D}_{\bar{I}}\bar{W}-3|W|) \, .
\eeq
Here $\{\Phi_I\}: I=1,\dots,n$ denote the $n$ complex components of the chiral superfields,
$D_I W = \partial_I W + K_I W$ and $K_I$ is the derivative of the field labeled by $I$~\footnote{We do
not consider here gauged supergravity models, nor models with nilpotent fields \cite{nil}}.
Equivalently, the action (\ref{comp_act}) can be rewritten in terms of the real and imaginary components of these complex fields:
\beq\label{components}
\Phi^I \; = \; \frac{1}{\sqrt{2}}(\chi^I + i \zeta^I) \, .
\eeq 
Substituting (\ref{components}) into the scalar potential (\ref{sugra_pot}), the action (\ref{comp_act}) takes the form
\begin{align}\label{re_act}
%
\notag S&= \int d^4x\, \sqrt{-g}\Bigg[\frac{1}{2}\,\left(\partial_{\mu}\chi^{I},\partial_{\mu}\zeta^I\right) \left(
\begin{matrix}
K_{I\bar{J}}^{\mathcal{R}} & K_{I\bar{J}}^{\mathcal{I}}\\
 - K_{I\bar{J}}^{\mathcal{I}} & K_{I\bar{J}}^{\mathcal{R}}
\end{matrix} 
\right) \left(
\begin{matrix}
\partial^{\mu}\chi^J\\
\partial^{\mu}\zeta^J
\end{matrix}
\right) - V(\boldsymbol{\chi},\boldsymbol{\zeta}) \Bigg]\\
&\equiv \int d^4x\, \sqrt{-g} \left[\frac{1}{2}G_{ij}\partial_{\mu}\phi^i\partial^{\mu}\phi^j - V(\boldsymbol{\phi})\right] \, ,
\end{align}
where $i=1,\dots,2n$, and $K_{I\bar{J}}^{\mathcal{R,I}}$ denote the real and imaginary parts of $K_{I\bar{J}}$. It can readily be verified that the real field metric $G_{ij}$ is symmetric. 
Equation (\ref{re_act}) will be the starting point for our review of multi-field inflation.

Assuming a spatially flat Friedmann-Robertson-Walker geometry, with the metric
\beq
ds^2 \; = \; dt^2 - a(t)^2\,d\bf{x}^2\ , 
\eeq
the classical (background) equations of motion for the spatially uniform real scalar fields and the scale factor correspond to
\beq\label{bkg1}
\ddot{\phi}^i+\Gamma^{i}_{jk}\dot{\phi}^j\dot{\phi}^k + 3H\dot{\phi}^i + G^{ij}V_{,j} \; = \; 0\, ,
\eeq
where
\beq \label{bkg2}
H^2 \; = \; \frac{1}{3}\left[\frac{1}{2}G_{ij}\dot{\phi}^i\dot{\phi}^j + V\right]\, ,
\eeq
and
\beq \label{bkg3}
\dot{H} \; = \; -\frac{1}{2}G_{ij}\dot{\phi}^i\dot{\phi}^j \, .
\eeq
Here $H \equiv \dot{a}/a$ denotes the Hubble parameter, $G^{ij}$ is the inverse field metric,
and $\Gamma^i_{jk}$ is the connection in field space.

Next, we generalize the treatment of  linear perturbations in \cite{Lalak:2007vi} 
in a way applicable to the model defined by the action (\ref{re_act}). 
The Newtonian (or longitudinal) gauge for the gravitational perturbations is a convenient gauge choice, 
since the anisotropic stress for the scalar fields vanishes in the linear approximation. The perturbed metric then takes the form
\beq
ds^2 \; = \; (1+2\Psi)dt^2 - a^2(1-2\Psi)d{\bf x}^2 \, .
\eeq
Decomposing the fields into the uniform background and the space-time-dependent perturbation
\beq
\phi^i(t,{\bf x}) \; = \; \phi^i (t) + \delta\phi^i(t,{\bf x})\, ,
\eeq
the resulting perturbed equations of motion for the fields in Fourier space read
\beq\label{per1}
\ddot{\delta\phi^i} + 2\Gamma^i_{jk}\dot{\phi}^j\dot{\delta\phi^k} + 3H\dot{\delta\phi^i} + \frac{k^2}{a^2}\delta\phi^i + \left[( G^{ij}V_{,j})_{,k}+\Gamma^i_{jl,k}\dot{\phi}^j\dot{\phi}^l\right]\delta\phi^k \; = \; 4\dot{\phi}^i\dot{\Psi} - 2G^{ik}V_{,k}\Psi
\eeq
while Einstein's equations lead to
\beq\label{per2}
\ddot{\Psi}+4H\dot{\Psi} + (\dot{H}+3H^2)\Psi \; = \; \frac{1}{2}\left[G_{ij}\dot{\phi}^i\dot{\delta\phi^j}+\frac{1}{2}G_{ij,k}\dot{\phi}^i\dot{\phi}^j\delta\phi^k-V_{,k}\delta\phi^k\right]\, ,
\eeq
\beq\label{per3}
3H(\dot{\Psi}+H\Psi)+ \dot{H}\Psi + \frac{k^2}{a^2}\Psi \; = \; -\frac{1}{2}\left[G_{ij}\dot{\phi}^i\dot{\delta\phi^j}+\frac{1}{2}G_{ij,k}\dot{\phi}^i\dot{\phi}^j\delta\phi^k+V_{,k}\delta\phi^k\right]\, ,
\eeq
\beq\label{per4}
H\Psi + \dot{\Psi} \; = \; \frac{1}{2}G_{ij}\dot{\phi}\delta\phi^j\, .
\eeq
In terms of the gauge-invariant Mukhanov-Sasaki variables,
\beq
Q^i \; \equiv \; \delta\phi^i + \frac{\dot{\phi}^i}{H}\Psi\, , 
\eeq
the multi-field perturbation equations (\ref{per1})-(\ref{per4}) reduce to
\beq\label{Q_is}
\ddot{Q}^i + 2\Gamma^i_{jk}\dot{\phi}^j\dot{Q}^k + 3H\dot{Q}^i + \frac{k^2}{a^2} Q^i + C^i_kQ^k \; = \; 0\, ,
\eeq
where the coefficients $C^i_k$ are defined as
\beq
\begin{aligned}
C^i_k \; \equiv \; &3G_{jk}\dot{\phi}^i\dot{\phi}^j - \frac{1}{2H^2}G_{jk}G_{lm}\dot{\phi}^i\dot{\phi}^j\dot{\phi}^l\dot{\phi}^m + \Gamma^i_{jl,k}\dot{\phi}^j\dot{\phi}^l + G^{ij}G_{lk}\frac{\dot{\phi}^l}{H}V_{,j} + \frac{\dot{\phi}^i}{H}V_{,k}\\
&\ + (G^{ij}V_{,j})_{,k}\ . 
\end{aligned}
\eeq
The equations (\ref{Q_is}) form a closed system for describing the gauge-invariant perturbations $Q^i$.
These equations are in general coupled and do not allow an immediate interpretation of the perturbations in terms of
observables. Instead, it is customary to introduce a kinematical basis in which the corresponding components of the 
perturbations can be related directly to the curvature and isocurvature perturbations, 
which in turn lead to the observable amplitudes and power spectra. 

The transition to the kinematical basis in the general multi-field scenario is discussed in~\cite{Peterson:2011yt}.
Here we specialize to the two-field scenario, $i=1,2$. In this case, the kinematical basis consists of the instantaneous 
directions parallel and orthogonal to the background trajectory in field space.
The speed in field space is conventionally denoted by $\dot{\sigma}$, where
\beq
\dot{\sigma}^2\; = \; G_{ij}\dot{\phi}^i\dot{\phi}^j \, ,
\eeq
The parallel and orthogonal directions to the background trajectory can be parametrized by the unit vectors
\beq
e_{\sigma}^i\equiv e_{\parallel}^i \; = \; \frac{\dot{\phi}^i}{\dot{\sigma}}\ , \qquad e_s^i\equiv e_{\perp}^i \; = \; \tilde{G}^i_j\frac{\dot{\phi}^j}{\dot{\sigma}}\, ,
\eeq
where we have defined
\beq
\tilde{G}^i_j \; \equiv \; \frac{\epsilon^{ik}G_{kj}}{\sqrt{G}}\, ,
\eeq
with $\epsilon^{12}=1$. In this new basis, we introduce the adiabatic and isocurvature perturbations 
\beq
\delta \sigma \; = \; G_{ij}e^{i}_{\sigma}\delta\phi^j\ , \qquad \delta s \; = \; G_{ij}e^{i}_{s}\delta\phi^j \
\eeq
and the directional derivatives
\beq
V_{\sigma} \; = \; e_{\sigma}^iV_{,i}\ , \ \ V_{s} \; = \; e_{s}^iV_{,i}\ , \ \ V_{\sigma \sigma}\; = \; e_{\sigma}^i e_{\sigma}^jV_{,ij}\ , \ \ V_{\sigma s} \; = \; e_{\sigma}^i e_{s}^j V_{,ij}\ , \ \ V_{ss} \; = \; e_{s}^ie_{s}^j V_{,ij} \, .
\eeq
The background equations can now be written in this basis. The background isocurvature is constant, $\dot{s}\equiv G_{ij}e^{i}_s\dot{\phi}^j=0$, 
while the adiabatic homogeneous equation of motion corresponds to
\beq
\ddot{\sigma} + 3H\dot{\sigma} + V_{\sigma} \; = \; 0 \, .
\eeq
In turn, the equations of motion for the gauge-invariant perturbations
\beq
Q_{\sigma} \; = \; G_{ij}e^{i}_{\sigma}Q^j\ , \qquad Q_s \; = \; G_{ij}e^{i}_{s}Q^j\ ,
\eeq
correspond to
\begin{align}\label{Qsig_eq}
\ddot{Q}_{\sigma} + 3H\dot{Q}_{\sigma} + 2\frac{V_s}{\dot{\sigma}}\dot{Q}_s+ \left(\frac{k^2}{a^2}+C_{\sigma\sigma}\right) Q_{\sigma} + C_{\sigma s}Q_s &= \; 0,\\ \label{Qs_eq}
\ddot{Q}_s + 3H\dot{Q}_s - 2\frac{V_s}{\dot{\sigma}}\dot{Q}_{\sigma}+ \left(\frac{k^2}{a^2}+C_{ss}\right)Q_s + C_{s\sigma}Q_{\sigma} & = \; 0 \, ,
\end{align}
where the background-dependent coefficients are
\begin{align}
C_{\sigma\sigma} & = \; V_{\sigma\sigma}-\left(\frac{V_s}{\dot{\sigma}}\right)^2 + \frac{2\dot{\sigma}}{H}V_{\sigma}+ 3\dot{\sigma}^2 - \frac{\dot{\sigma}^4}{2H^2}+ \Gamma^l_{ik}G_{lj}\dot{\phi}^i\dot{\phi}^j\dot{\phi}^k\frac{V_{\sigma}}{\dot{\sigma}^3} + \epsilon_{il}\Gamma^l_{jk}\dot{\phi}^i\dot{\phi}^j\dot{\phi}^k\frac{V_s}{\dot{\sigma}^3}\,,\\
\notag C_{\sigma s} & = \; 6H\frac{V_s}{\dot{\sigma}}+2\frac{V_s V_{\sigma}}{\dot{\sigma}^2} + 2V_{\sigma s} + \frac{\dot{\sigma}V_s}{H} -2\tilde{G}^l_iG_{mk}\Gamma^m_{lj}\dot{\phi}^i\dot{\phi}^j\dot{\phi}^k\frac{V_{\sigma}}{\dot{\sigma}^3}\\
&\quad  - 2\tilde{G}^m_i\tilde{G}^l_jG_{nl}\Gamma^n_{km}\dot{\phi}^i\dot{\phi}^j\dot{\phi}^k\frac{V_s}{\dot{\sigma}^3}\,,\\
C_{s\sigma} &= \; - 6H\frac{V_s}{\dot{\sigma}} - 2\frac{V_s V_{\sigma}}{\dot{\sigma}^2} + \frac{\dot{\sigma}V_s}{H}\,,\\
\notag C_{ss} &= \; V_{ss} - \left(\frac{V_s}{\dot{\sigma}}\right)^2 - \tilde{G}^l_j\tilde{G}^m_kG_{in}\Gamma^n_{lm}\dot{\phi}^i\dot{\phi}^j\dot{\phi}^k\frac{V_{\sigma}}{\dot{\sigma}^3} + \tilde{G}^l_kG_{jm}\Gamma^m_{il}\dot{\phi}^i\dot{\phi}^j\dot{\phi}^k \frac{V_s}{\dot{\sigma}^3}\\
&\quad - \frac{1}{2}\left(\tilde{G}^k_i\tilde{G}^{mj}G_{ml}\Gamma^l_{kj} + \tilde{G}^k_i\Gamma^l_{kl}\right)\dot{\phi}^i\frac{V_s}{\dot{\sigma}} + \frac{1}{2}R\dot{\sigma}^2\ .
\end{align}
Here $R$ denotes the curvature scalar, $\epsilon_{12}=\sqrt{G}\epsilon^{12}$,
and $\tilde{G}^{ij}=G^{-1}\epsilon^{ik}\epsilon^{jl}G_{kl}$.
The adiabatic perturbation $Q_{\sigma}$ is related to the comoving curvature perturbation as:
\beq
\mathcal{R} \; = \; \frac{H}{\dot{\sigma}}Q_{\sigma} \, .
\eeq
The isocurvature perturbation in turn defines the entropy perturbation,
\beq
\mathcal{S} \; = \; \frac{H}{\dot{\sigma}}Q_s \, .
\eeq
At the start of inflation, deep inside the Hubble radius, the adiabatic and entropy fluctuations are initially statistically independent, 
and can be approximated by the corresponding Minkowski vacuum state:
\beq\label{init_vac}
Q_{\sigma}(\tau_i) \; = \; \frac{e^{-ik\tau_i}}{a(\tau_i)\sqrt{2k}}\xi_{\sigma}(k)\ , \qquad Q_s(\tau_i) \; = \; \frac{e^{-ik\tau_i}}{a(\tau_i)\sqrt{2k}}\xi_s(k)\,,
\eeq
where $d\tau\equiv dt/a$ is the conformal time, and $\xi_I: I=\sigma,s$ are independent Gaussian random variables satisfying
\beq
\langle \xi_I(k) \rangle \; = \; 0\ , \quad \langle \xi_I(k),\bar{\xi}_J(k') \rangle = \delta_{IJ}\delta(k-k') \, .
\eeq
Given the adiabatic and entropy perturbations, their power spectra are defined as
\begin{align}
\langle \mathcal{R}(k)\bar{\mathcal{R}}(k') \rangle &= \frac{2\pi^2}{k^3}\mathcal{P}_{\mathcal{R}}\,\delta(k-k'),\\
\langle \mathcal{S}(k)\bar{\mathcal{S}}(k') \rangle &= \frac{2\pi^2}{k^3}\mathcal{P}_{\mathcal{S}}\,\delta(k-k').
\end{align}
Equations (\ref{Qsig_eq}),(\ref{Qs_eq}) couple the adiabatic and entropy perturbations.
With the initial conditions given by (\ref{init_vac}), the curvature perturbation at the end of inflation will
be a linear combination of the adiabatic and entropy Gaussian random variables, 
$\mathcal{R} = \mathcal{R}_1 \xi_{\sigma}+\mathcal{R}_2\xi_s$. In terms of these components, the power spectrum takes the form
\beq\label{PR}
\mathcal{P}_{\mathcal{R}} \; = \; \frac{k^3}{2\pi^2}\left(|\mathcal{R}_1|^2+|\mathcal{R}_2|^2\right).
\eeq
The same argument applies to the entropy power spectrum, $\mathcal{S} = \mathcal{S}_1 \xi_{\sigma}+\mathcal{S}_2\xi_s$ and 
\beq\label{PS}
\mathcal{P}_{\mathcal{S}} \; = \; \frac{k^3}{2\pi^2}\left(|\mathcal{S}_1|^2+|\mathcal{S}_2|^2\right).
\eeq
Equations (\ref{Qsig_eq}) and (\ref{Qs_eq}) can be integrated numerically. 

To avoid the introduction of random variables, the statistical independence of the adiabatic and 
entropy perturbations is taken into account by integrating the equations twice: 
first with $Q_{\sigma}$ initially equal to the corresponding Minkowski value and $Q_s=0$, 
then with $Q_s$ equal to the Minkowski vacuum value and $Q_{\sigma}=0$. 
The end of inflation is chosen to correspond to the end of slow roll, $w\equiv P/\rho = -1/3$. 
The resulting power spectra for the curvature and entropy perturbations are then calculated as in (\ref{PR}) and (\ref{PS}), 
where in this case the indices $1,2$ denote the corresponding numerical run. 
If the numerical code were to be run only once with both $Q_{\sigma}$ and $Q_s$ in the vacuum state,
the resulting curvature perturbation would be given by $\mathcal{R}=\mathcal{R}_1+\mathcal{R}_2$, 
with the power spectrum $\mathcal{P}_{\mathcal{R}}=\frac{k^3}{2\pi^2}|\mathcal{R}_1+\mathcal{R}_2|^2$,
which is different from (\ref{PR}) \cite{Tsujikawa:2003}.
Finally, the observables, the scalar tilt $n_s$ and the tensor to scalar ratio $r$, are derived from the power spectrum
using their definitions:
\beq
n_s \; = \; 1 + \frac{d\log\mathcal{P}_{\mathcal{R}}}{d\log k}\ , \qquad r \; = \; \frac{\mathcal{P}_{T}}{\mathcal{P}_{\mathcal{R}}}\,,
\eeq
where $\mathcal{P}_{T}$ denotes the spectrum of the tensor perturbations. It has the same form as in the single-field limit, 
$\mathcal{P}_{T}=\left.\frac{2}{\pi^2}H^2\right|_{k=aH}$, since at linear order the scalar perturbations decouple from
the vector and tensor perturbations~\cite{tensorrefs}.

In the particular case when the theory contains a single chiral superfield $\Phi=(\chi+i\zeta)/\sqrt{2}$, or when the dynamics of $n-1$ of the complex fields is constrained so that they can be replaced by their expectation values, the metric $G_{ij}$ reduces to
\beq
(G_{ij}) \; = \; K_{\Phi\bar{\Phi}}^{\mathcal{R}}\left(\begin{matrix}1 & 0\\ 0 & 1\end{matrix}\right) \equiv f(\chi,\zeta)\left(\begin{matrix}1 & 0\\ 0 & 1\end{matrix}\right)\ .
\eeq
In this case, the background equations (\ref{bkg1})-(\ref{bkg3}) take the form
\begin{align}\label{bkg_sim}
&\ddot{\chi}+3H\dot{\chi}+\frac{1}{2}f^{-1}\left[(\dot{\chi}^2-\dot{\zeta}^2)f_{\chi} + 2\dot{\chi}\dot{\zeta}f_{\zeta}\right] + f^{-1}V_{,\chi} \; = \; 0\\
&\ddot{\zeta}+3H\dot{\zeta}+\frac{1}{2}f^{-1}\left[(\dot{\zeta}^2-\dot{\chi}^2)f_{\zeta} + 2\dot{\chi}\dot{\zeta}f_{\chi}\right] + f^{-1}V_{,\zeta} \; = \; 0\\
& H^2 \; = \; \frac{1}{3}\left[\frac{1}{2}f(\dot{\chi}^2+\dot{\zeta}^2) + V\right]\\
& \dot{H} \; = \; -\frac{1}{2}f(\dot{\chi}^2+\dot{\zeta}^2)
\end{align}
while the perturbation equations correspond to (\ref{Qsig_eq}),(\ref{Qs_eq}), where
\begin{align} \label{simpcoef1}
C_{\sigma\sigma} &= \; V_{\sigma\sigma} - \left(\frac{V_s}{\dot{\sigma}}\right)^2 + \frac{2\dot{\sigma}}{H}V_{\sigma} + 3\dot{\sigma}^2 - \frac{\dot{\sigma}^4}{2H^2} - \frac{f_{\chi}\dot{\chi} + f_{\zeta}\dot{\zeta}}{2f\dot{\sigma}}V_{\sigma} - \frac{f_{\zeta}\dot{\chi}-f_{\chi}\dot{\zeta}}{2f\dot{\sigma}}V_s,\\ \label{simpcoef2}
C_{\sigma s} &= \; 6H\frac{V_s}{\dot{\sigma}} + 2\frac{V_s V_{\sigma}}{\dot{\sigma}^2} + 2V_{\sigma s} + \frac{\dot{\sigma}V_s}{H} - \frac{f_{\chi}\dot{\chi}+f_{\zeta}\dot{\zeta}}{f\dot{\sigma}}V_s + \frac{f_{\zeta}\dot{\chi}-f_{\chi}\dot{\zeta}}{f\dot{\sigma}}V_{\sigma},\\ \label{simpcoef3}
C_{s\sigma} &= \; -6H\frac{V_s}{\dot{\sigma}} - 2\frac{V_{\sigma}V_s}{\dot{\sigma}^2} + \frac{\dot{\sigma}V_s}{H},\\ \label{simpcoef4}
C_{ss} &= \; V_{ss}-\left(\frac{V_s}{\dot{\sigma}}\right)^2 + \frac{f_{\chi}\dot{\chi}+f_{\zeta}\dot{\zeta}}{2f\dot{\sigma}}V_{\sigma} + \frac{f_{\zeta}\dot{\chi}-f_{\chi}\dot{\zeta}}{2f\dot{\sigma}}V_{s} + \frac{\dot{\sigma}^2}{2f^2}\left(\frac{f_{\chi}^2}{f} + \frac{f_{\zeta}^2}{f} - f_{\chi\chi} - f_{\zeta\zeta}\right) \, ,
\end{align}
with ${\bf e}_{\sigma}=(\dot{\chi}/\dot{\sigma},\dot{\zeta}/\dot{\sigma})$ and
${\bf e}_s=(\dot{\zeta}/\dot{\sigma},-\dot{\chi}/\dot{\sigma})$. In this particular scenario, 
the perturbation equations can be brought into an integral form in the slow roll approximation~\cite{mukhanov:1998}.
Nevertheless, equations (\ref{Qsig_eq}, \ref{Qs_eq}) with coefficients (\ref{simpcoef1})-(\ref{simpcoef4}) can
readily be solved numerically in general, even for non-trivial forms of the effective sigma-model function $f(\chi,\zeta)$, 
which are commonly found in strongly-stabilized models in minimal and no-scale supergravity.

\section{Two-Field Analysis of the No-Scale Model}

In our previous paper~\cite{EGNO2}, we constrained the motion of the inflaton field in $(\rho, \alpha)$ space
(or, equivalently, in $({\cal R}e\,T,{\cal I}m\,T)$ space) by modifying the K\"ahler potential so that the
$T$ field is stabilized. Such stabilization is a necessary feature of any string scenario, and might arise from
corrections to the minimal no-scale form induced by either perturbative or non-perturbative effects,
though the general forms of such corrections are unknown. Here we consider the following modified K\"ahler potential:
\beq\label{ktheta}
K \; = \; - \, 3\log\left(T+\bar{T} - c\left[\cos\theta(T+\bar{T}-1)-\sin\theta(T-\bar{T})^2\right]^2 \right)+\frac{|\phi|^2}{(T+\bar{T})^3} \, ,
\eeq
where $\theta$ and $c$ are free constant parameters. It is apparent that the corrections are quadratic for
$\theta = 0$ and quartic for $\theta = \pi/2$, with a combination at intermediate $\theta$.
For values of $c$ that are sufficiently large, the inflaton is confined to a narrow ``bobsleigh track", and its dynamics
reduces to that of a single real field. 

In~\cite{EGNO2} we considered a purely quartic modification of the no-scale K\"ahler potential
with $c = 1000$ as a default option, and showed
that for varying $\theta \in [0, \pi/2]$ the predictions of the model interpolated between those
of chaotic quadratic inflation at $\theta = 0$ and the Starobinsky model at $\theta = \pi/2$.
Here we explore the new possibilities opened up by (\ref{ktheta}) for smaller values of $c$, finding that
two-field effects open up new options
for interpolating between BICEP2- and Planck-friendly predictions for $r$, even for $\theta = 0$, while retaining the previous
model's successful predictions for $n_s$ and exhibiting small values of the non-Gaussianity parameter $f_{\rm NL}$.

\subsection{`Imaginary' Inflation}

We consider initial conditions $\phi=0$ as enforced by (\ref{ourW}) when $a = 1/2$, and
consider first the possibility that $\theta=0$.
In this case, ${\cal R}e~T$ (or, equivalently, $\rho$) is constrained by a quadratic potential term,
and the effective Lagrangian along the imaginary direction has the form
\beq
\mathcal{L} \; = \; \frac{1}{2}(1+2c)\partial_{\mu}\alpha\partial^{\mu}\alpha - \frac{1}{2}m^2\alpha^2 \, ,
\eeq
which corresponds to the usual quadratic chaotic Lagrangian in terms of the canonically-normalized field
\beq
A \; \equiv \; \alpha (1+2c)^{1/2}
\label{canon}
\eeq
with the physical mass
\beq
M \; = \;  \frac{m}{(1+2c)^{1/2}}  \, .
\eeq
Naively, one might expect that this model would necessarily yield a large value of $r$,
as in minimal quadratic chaotic inflation. However,
even with $\rho = 0$ initially, the full two-field analysis shows that the inflaton trajectory
evolves through non-zero values of $\rho$ when $c$ is not very large. Moreover, we
find that when $c$ is not large feed-through from isocurvature perturbations to curvature perturbations is capable
of enhancing considerably the scalar perturbation spectrum, and hence suppressing $r$.

\begin{figure}[!h]
\centering
	\scalebox{0.8}{\includegraphics{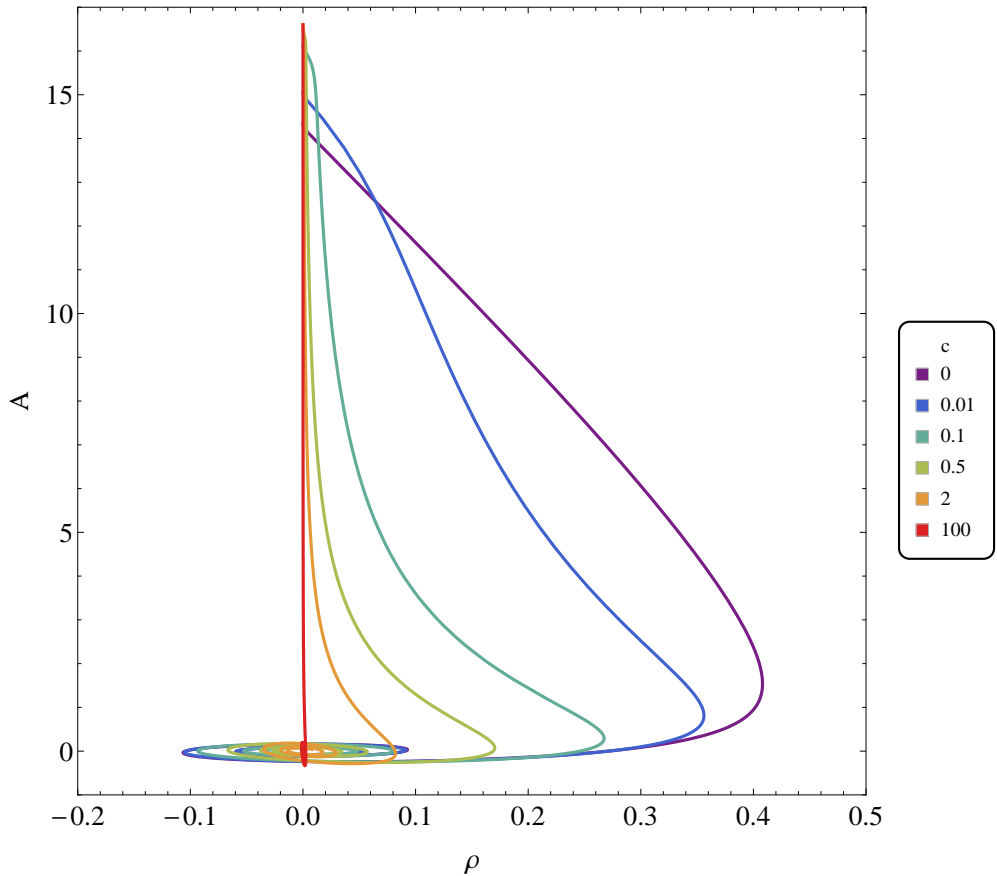}} 
	\caption{\it Inflationary trajectories for the K\"ahler potential (\ref{ktheta}) at $\theta=0$ for selected
	colour-coded values of $c$.
	The initial conditions correspond to $\rho=0$ and $A$ tuned to yield 70 e-folds of inflation.} \label{n2traj}
\end{figure}

The inflaton evolution in $(\rho,A)$ field space for selected colour-coded values of $c$
is displayed in Fig.~\ref{n2traj}. The initial condition is taken along the imaginary axis in the inflaton field space,
i.e., $\rho =0$ initially, and the initial value of $A$ is chosen to yield $N = 70$ e-folds of inflation. We see that
the trajectory remains very close to the ${\cal I}m~T$ axis when $c = 100$, apart from small deviations in the final
`circling the drain' phase of the evolution. The deviations in $\rho$ are very visible for the choices $c = 2$ and 0.5
also shown in Fig.~\ref{n2traj}, but  the deviations in these cases are also not significant during the inflationary
phase. It is only for $c \la 0.1$ that the inflaton trajectory is modified significantly during the inflationary phase.
We see that there is a large deviation to $\rho \ne 0$ throughout the inflationary phase when $c = 0.01$, which is
qualitatively similar to the case $c = 0$ when the modification to the K\"ahler structure in (\ref{ktheta}) is removed.

\begin{figure}[h!]
\centering
	\scalebox{0.9}{\includegraphics{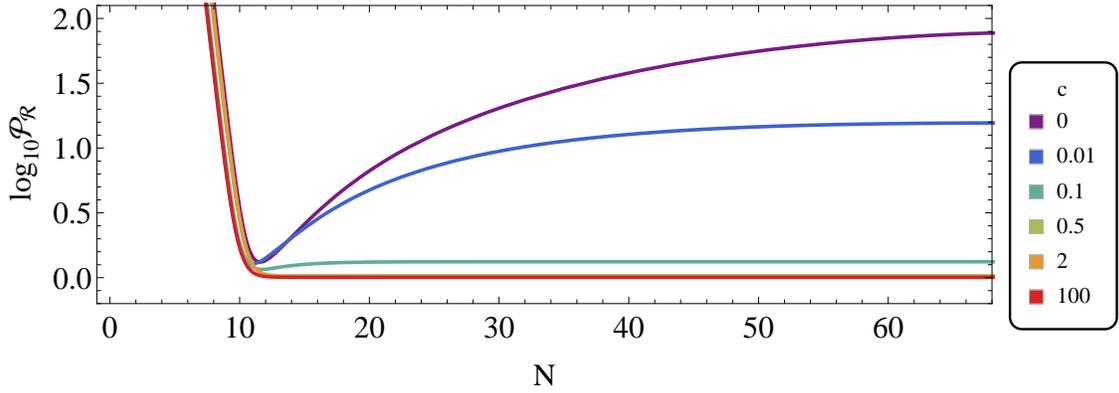}} 
	\caption{\it The curvature power spectrum $\mathcal{P}_{\mathcal{R}}$ for the trajectories shown in
	Fig.~\protect\ref{n2traj} corresponding to different colour-coded values of $c$.
	The curves are normalized relative to the single-field expression, 
	and correspond to the mode that leaves the horizon at the start of the last 60 e-folds of inflation.} \label{n2ps}
\end{figure}

The enhancement of the curvature power spectrum as function of the number
of e-folds before the end of inflation, $N$, is displayed in Fig.~\ref{n2ps}, for the same colour-coded
choices of $c$ as in Fig.~\ref{n2traj}. Here $\mathcal{P}_{\mathcal{R}}$ is normalized to the single-field expression,
and corresponds to the mode that leaves the horizon at the start of the last 60 e-folds of inflation.
We see that there is no significant enhancement for $c \ge 0.5$, and that the enhancement is not large for $c = 0.1$.
However, the curvature power spectrum is enhanced by an order of magnitude for $c =  0.01$,
and by almost two orders of magnitude for $c = 0$.

\begin{figure}[!h]
\centering
	\scalebox{0.9}{\includegraphics{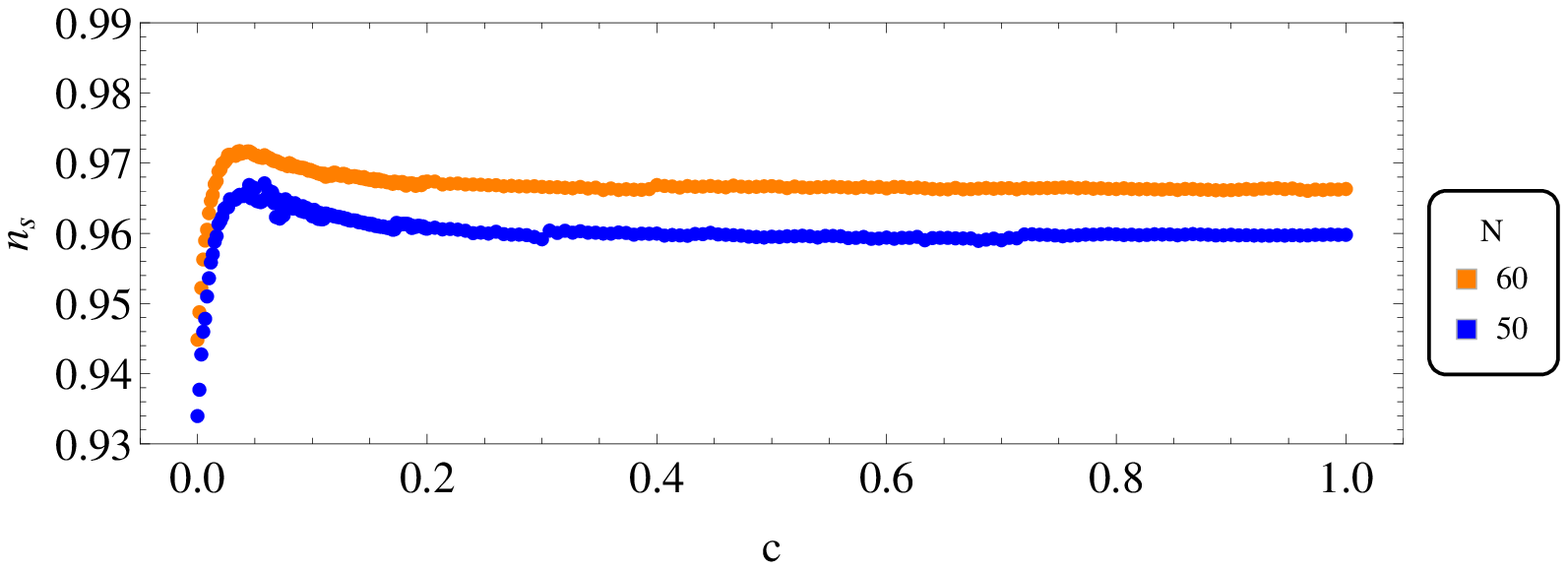}} 
	\vspace{10pt}
	\scalebox{0.9}{\includegraphics{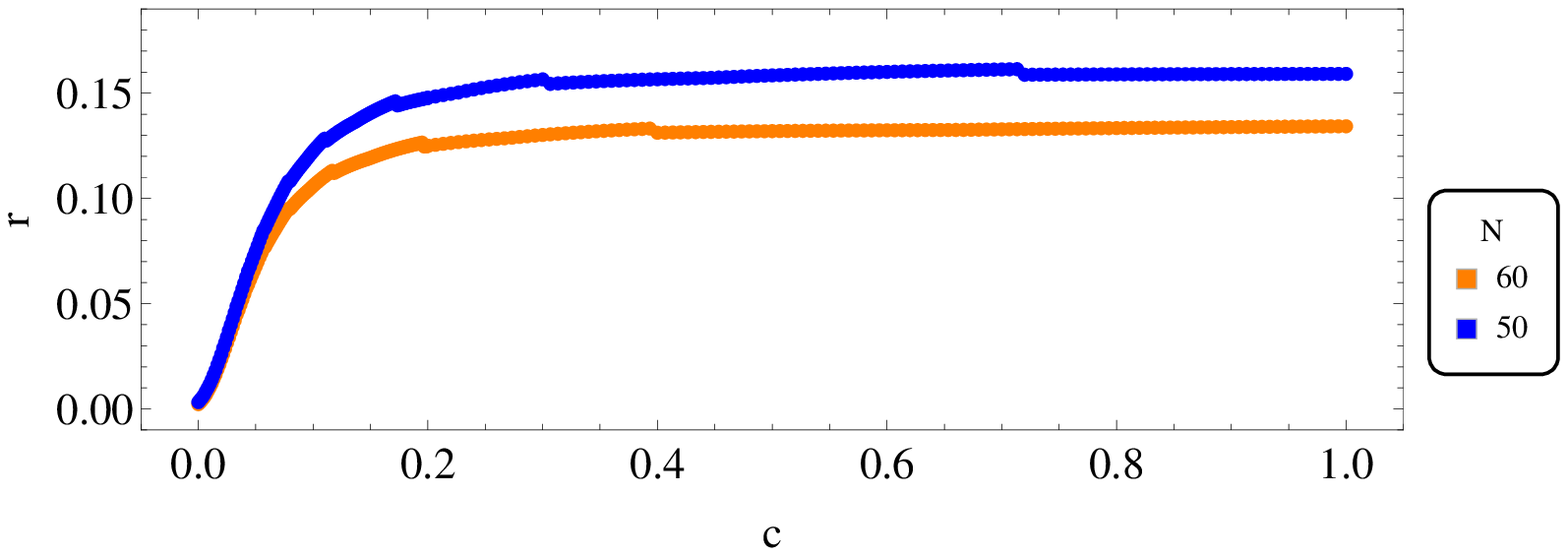}} 
	\caption{\it The dependences on the parameter $c$ of the scalar spectral index $n_s$ (upper panel)
	and the tensor-to-scalar ratio $r$ (lower panel), assuming either $N = 50$ or $60$ e-folds
	(blue and orange lines, respectively) and ${\cal R}e~T = 0$
	initially, i.e., $\theta=0$.} \label{n2nsr1}
\end{figure}

As a consequence of the enhancement in $\mathcal{P}_{\mathcal{R}}$,
the scalar tilt $n_s$ and the tensor-to-scalar ratio $r$ are reduced relative to their values in
the single-field chaotic scenario with a quadratic potential, as seen in Fig.~\ref{n2nsr1}. 
In particular, compatibility with the range of $n_s$ allowed by Planck is lost for $c\lesssim 0.05$,
as seen in the upper panel of Fig.~\ref{n2nsr1}. The lower panel of Fig.~\ref{n2nsr1} shows that
$r$ has a plateau with a value similar to that in the chaotic quadratic single-field model
when the stabilizing parameter $c \ga 0.2$, falling to much
smaller values as $c \to 0$. Therefore, two-field effects provide a mechanism for suppressing
$r$ in this type of quadratic inflationary model, that is complementary to varying the angle of the
``bobsleigh track" as discussed in~\cite{EGNO2}.

\begin{figure}[!h]
\centering
	\scalebox{0.95}{\includegraphics{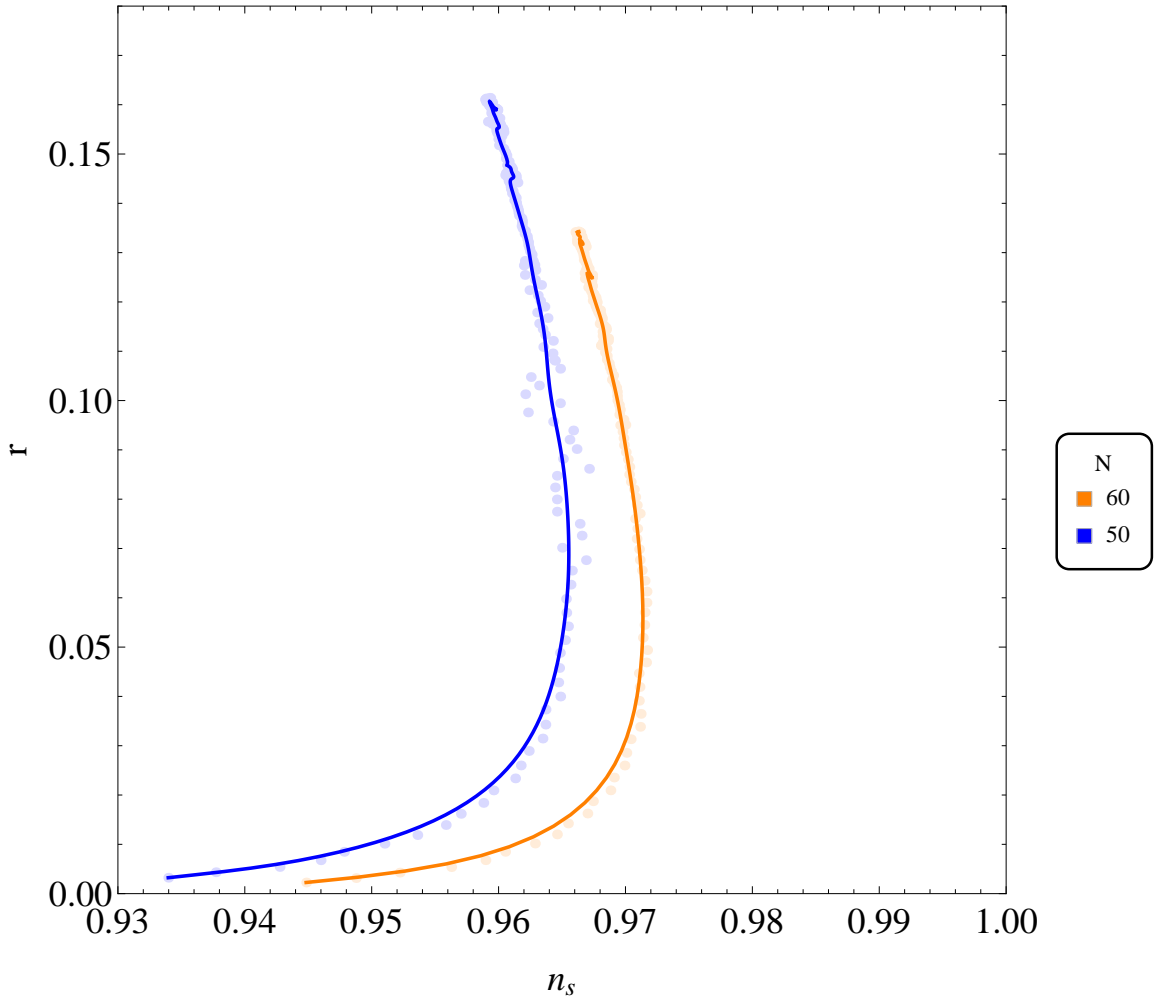}} 
	\caption{\it Parametric curve $(n_s(c),r(c))$, $c \in (0,1)$ 
	assuming initial conditions with ${\cal R}e~T = 0$,
i.e., $\theta=0$, assuming either $N = 50$ or $60$ e-folds (blue and orange lines, respectively).} \label{n2nsr2}
\end{figure}

The enhancement of the curvature power spectrum is completely negligible for $c\ga1$, 
where the single-field expressions can be trusted. Therefore, pure quadratic inflation is recovered in the limit of large $c$.
Fig.~\ref{n2nsr2} shows the parametric curve $(n_s(c),r(c))$ for $c$ decreasing from $1 \to 0$ from top to
bottom of the plot. This shows that, even with initial conditions
along the imaginary direction in the $T$ plane, one can interpolate between BICEP2- and Planck-friendly
values of $r \ga 0.1, \la 0.1$, respectively, by varying the stabilization parameter $c$.
We recall that the modification of the K\"ahler potential in (\ref{ktheta}) has no deep theoretical
justification. {\it A priori}, one might have expected any such modifications of the simple no-scale form
(\ref{minimal}) to have coefficients that are ${\cal O}(1)$ in natural units, so it is interesting that the
naive single-field results would be recovered in this case. This result could perhaps have been expected, since
the energy scale during inflation is very small in natural units. We recall that in~\cite{EGNO2} our default choice was
$c = 1000$, for which the single-field approximation used there is amply justified. Finally, we note an important
feature of Fig.~\ref{n2nsr2} in the limit of small $c$: although $r$ may become as small as in the Starobinsky
model, this is possible only for unacceptably small values of $n_s$ when an initial condition with ${\cal R}e~T = 0$,
i.e., $\theta = 0$, is chosen.

\subsection{`Real' Inflation}

\begin{figure}[!h]
\centering
	\scalebox{0.9}{\includegraphics{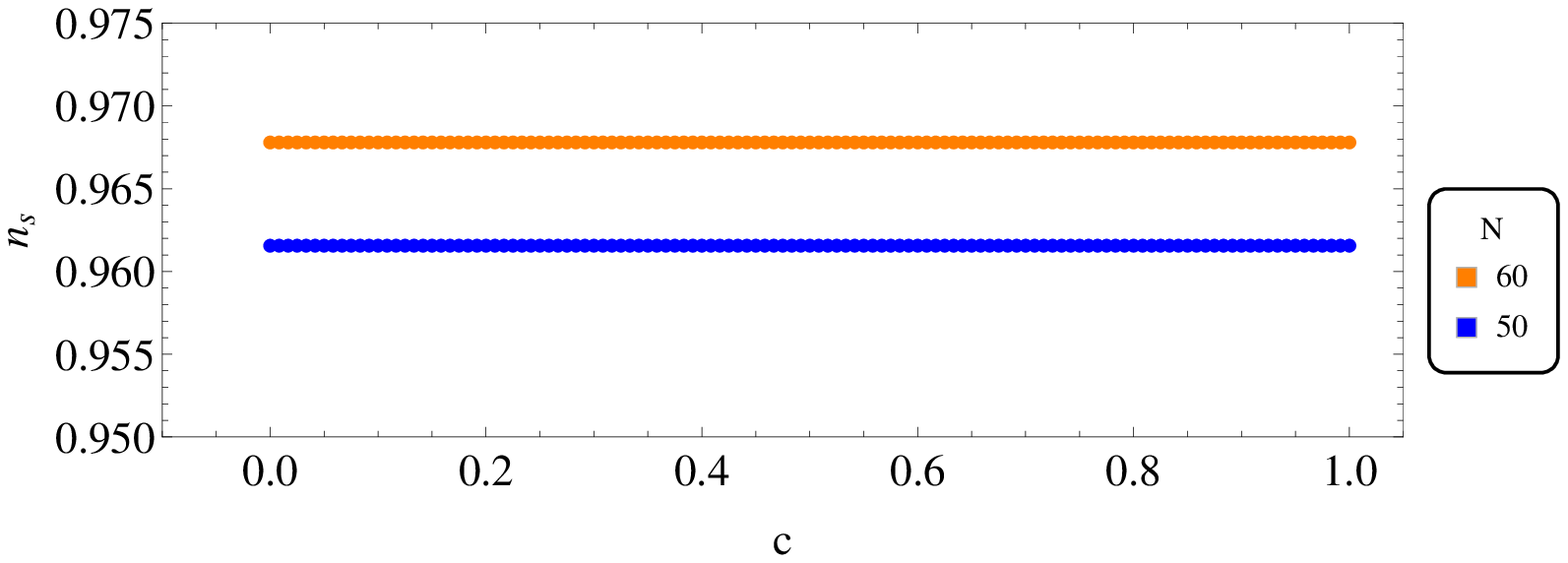}} 
	\scalebox{0.9}{\includegraphics{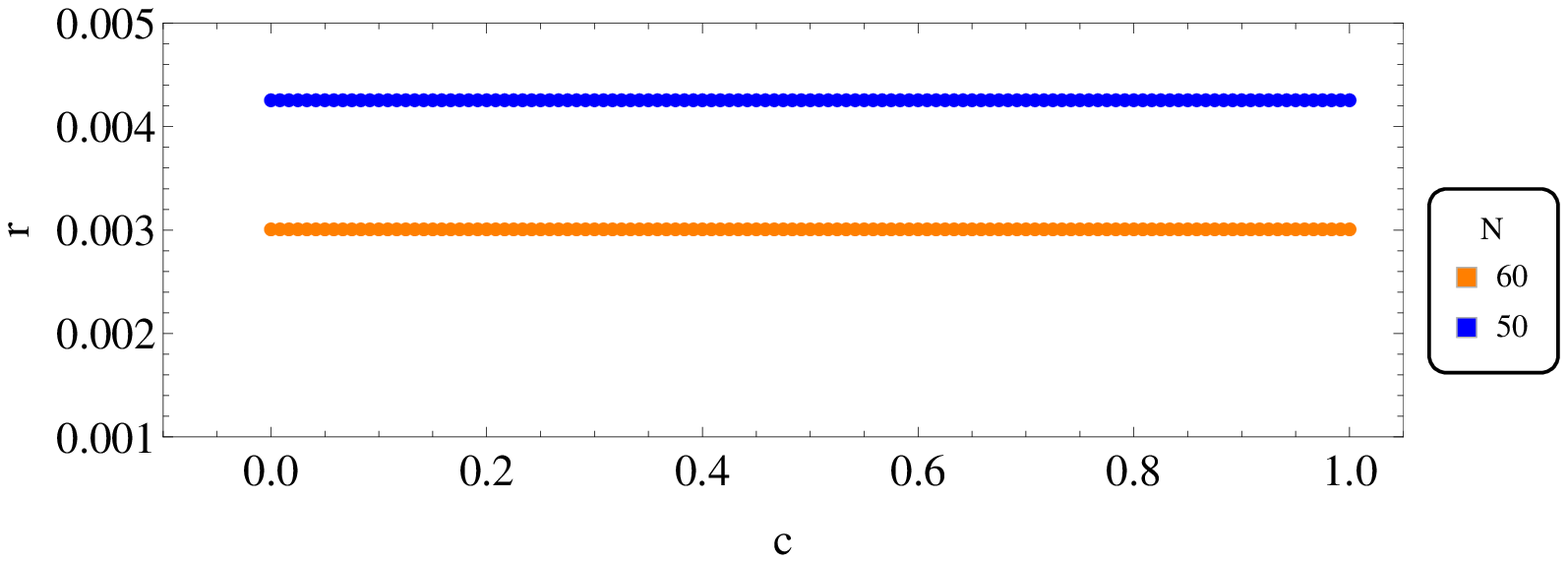}} 
	\caption{\it The dependences on the parameter $c$ of the scalar spectral index $n_s$ (upper panel)
	and the tensor-to-scalar ratio $r$ (lower panel), assuming either $N = 50$ or $60$ e-folds 
	(blue and orange lines, respectively) assuming ${\cal I}m~T = 0$
	initially, i.e., $\theta=\pi/2$.} \label{n2nsr_re}
\end{figure}

We now consider inflation along the direction where ${\cal I}m~T = 0$, i.e., $\alpha = 0$ and $\theta=\pi/2$,
with the inflaton identified as ${\cal R}e~T$, i.e., $\rho$. We recall that when
$\theta = \pi/2$ the imaginary direction is quartically constrained. For this reason, there is no
renormalization of any kinetic or mass term as occurred for `imaginary' inflation. In the case of `real' inflation, as we
see in Fig.~\ref{n2nsr_re}, the values of $n_s$ and $r$ are essentially identical for any $c \ne 0$.
As could be expected, in this direction the no-scale supergravity model has the same
predictions as the Starobinsky model, which are completely compatible with the Planck data.

\subsection{`Complex' Inflation}

We showed previously~\cite{EGNO2} that single-field models interpolating between BICEP2- friendly
chaotic quadratic inflation and the Planck-friendly Starobinsky model could be obtained by
varying $\theta$ in (\ref{ktheta}), keeping the stabilization coefficient $c$ large, and we showed
in Section~3.1 that two-field effects could also interpolate between the quadratic and Starobinsky
predictions even when $\theta = 0$. We now
explore the possibilities for $0 < \theta < \pi/2$ and varying $c$, finding that
BICEP2- and/or Planck-compatible predictions for general $\theta>0$ even
when the stabilization coefficient $c < 1$.

\begin{figure}[!h]
\centering
	\scalebox{0.49}{\includegraphics{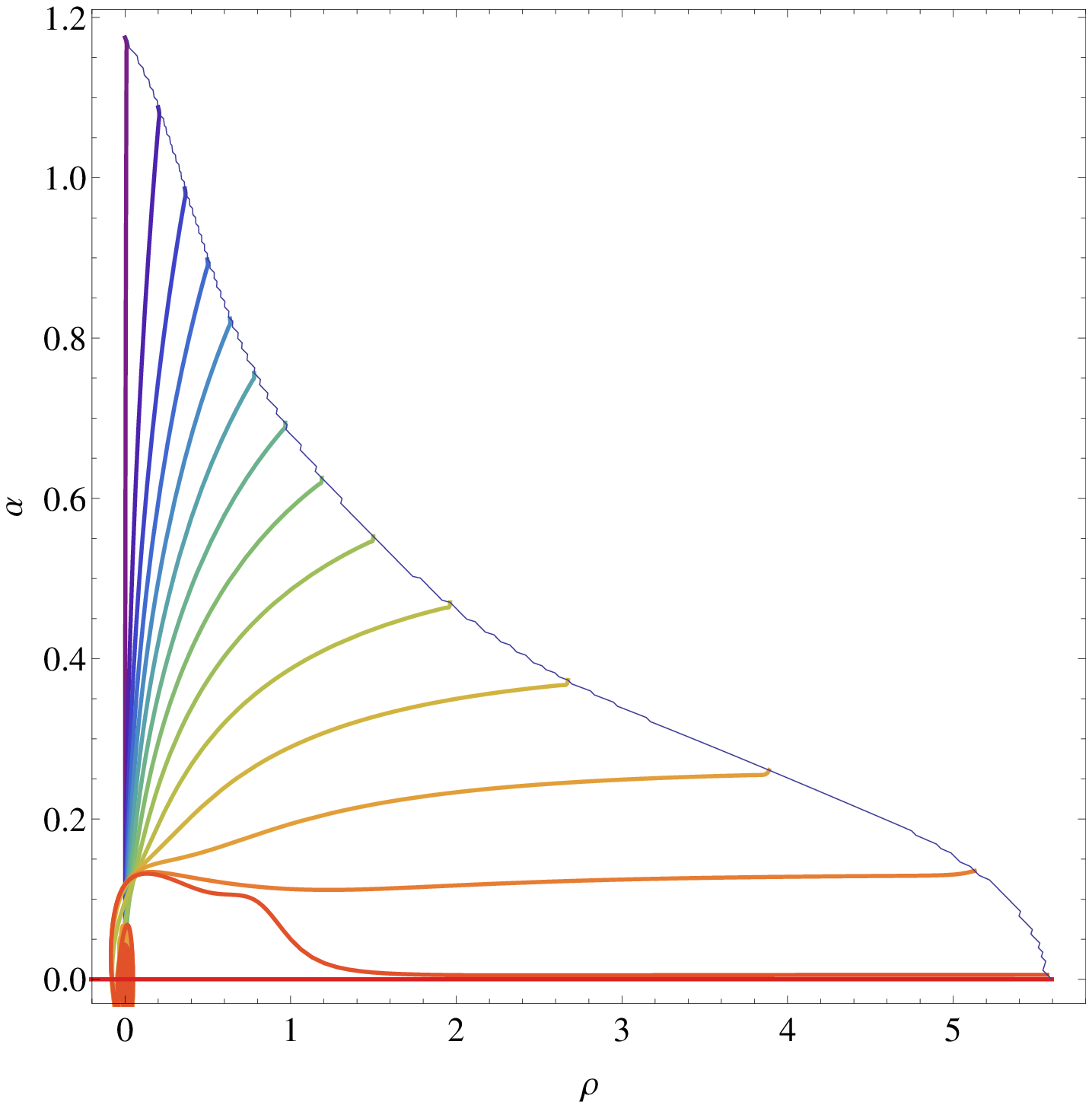}} 
	\scalebox{0.49}{\includegraphics{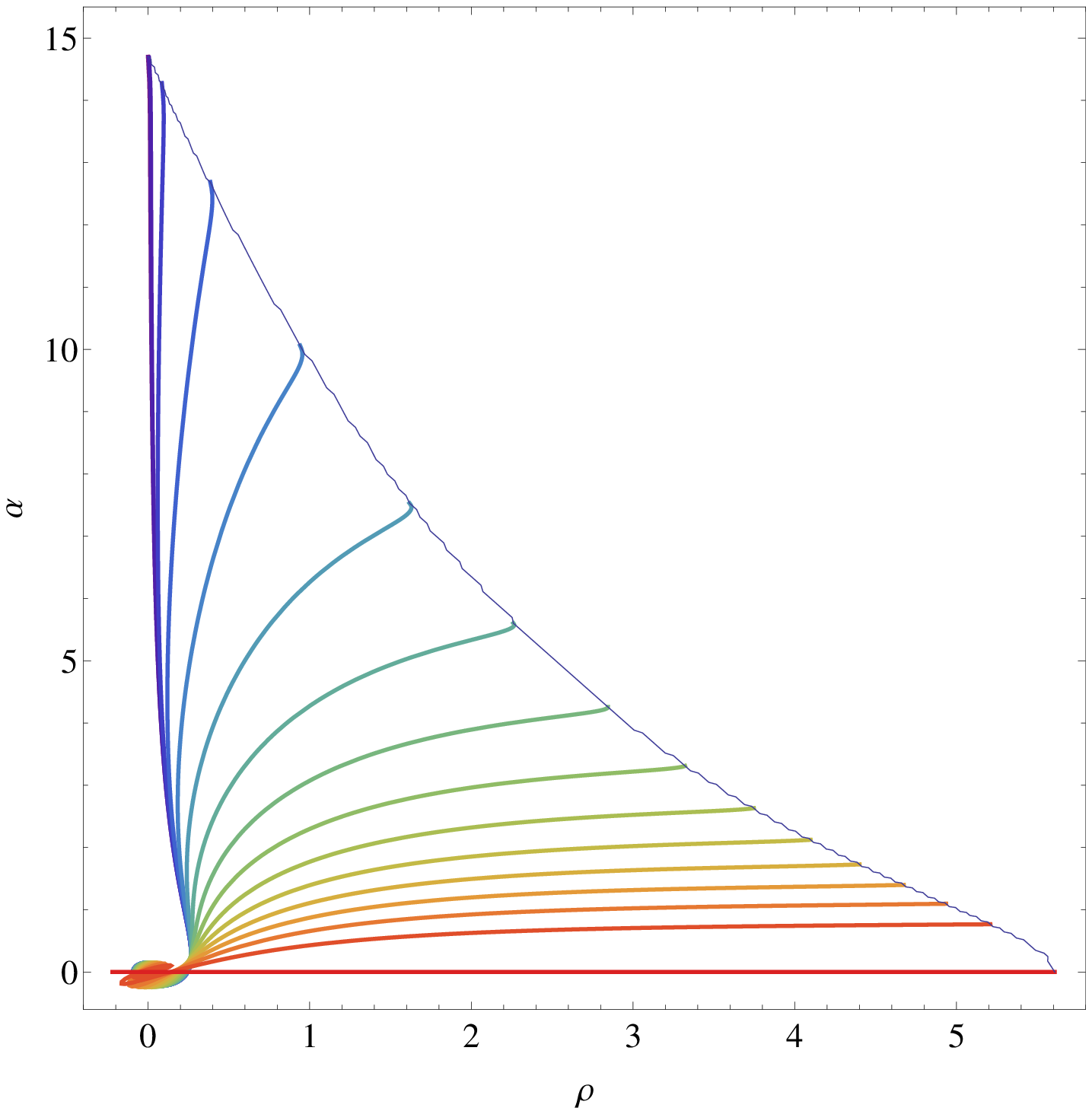}} 
	\caption{\it Inflationary trajectories for the K\"ahler potential (\ref{ktheta})
	at $c=100$ (left panel) and $c = 0.1$ (right panel). In both cases, the initial conditions are
	chosen so as to yield 70 e-folds of inflation. The pure Starobinsky limit is recovered along the real direction
	${\cal I}m~T (\alpha) = 0$.} \label{thtraj}
\end{figure}

Fig.~\ref{thtraj} shows a selection of inflationary trajectories in the $(\rho,\alpha)$ plane for $c=100$
(left panel) and $c = 0.1$ (right panel). All the trajectories start from the $N_{\rm tot}=70$ contour line,
where $N$ denotes the number of e-folds. In the $c = 100$ case, the trajectories are generally
convex until they start `circling the drain',  with the exception of $\theta=\pi/2$, 
for which the evolution is strictly along the real axis. However, one sees
deviations in the evolution even for angles very close to the real axis. For example,
we show in Fig.~\ref{thtraj} the trajectory for $\theta = \pi/2 (1-\cos(\pi/600))$, which 
exhibits a substantial departure from the axis at $\rho \approx 1.4$.
In the $c = 0.1$ case, the trajectories all become concave for small $\rho$ and $\alpha$, and
`circle the drain' clockwise.

\begin{figure}[!h]
\centering
	\scalebox{0.9}{\includegraphics{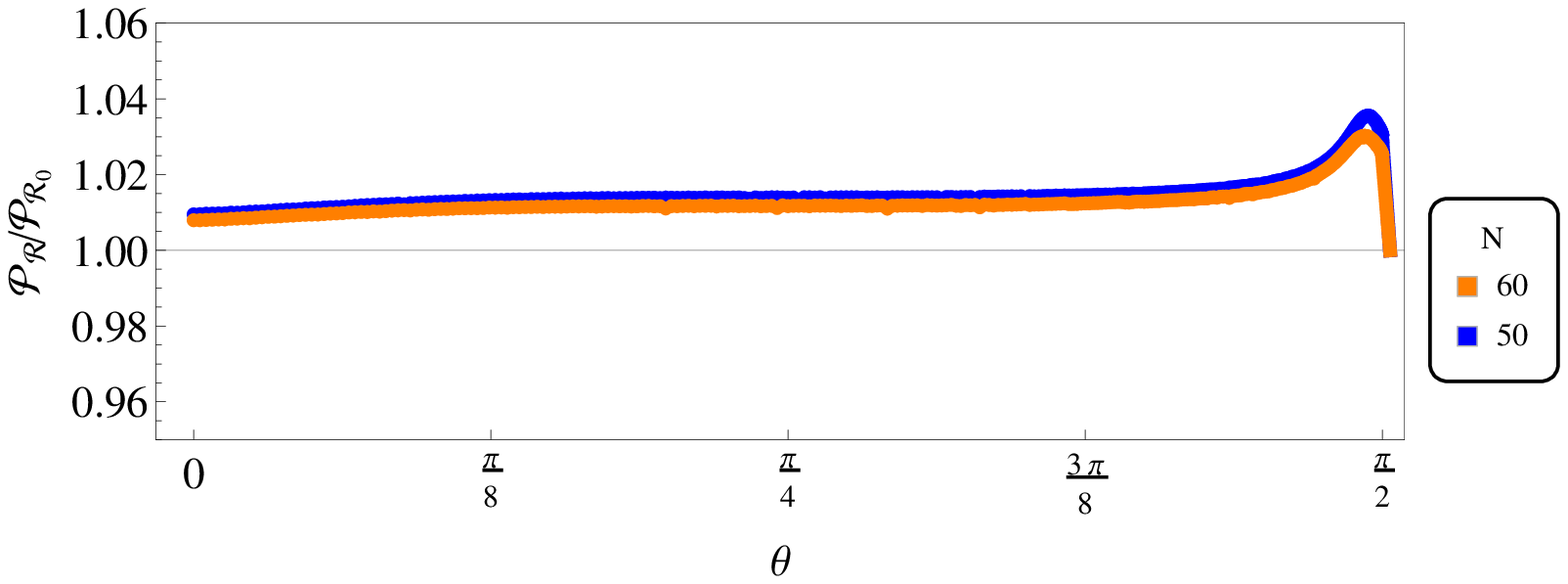}} \\
	\scalebox{0.9}{\includegraphics{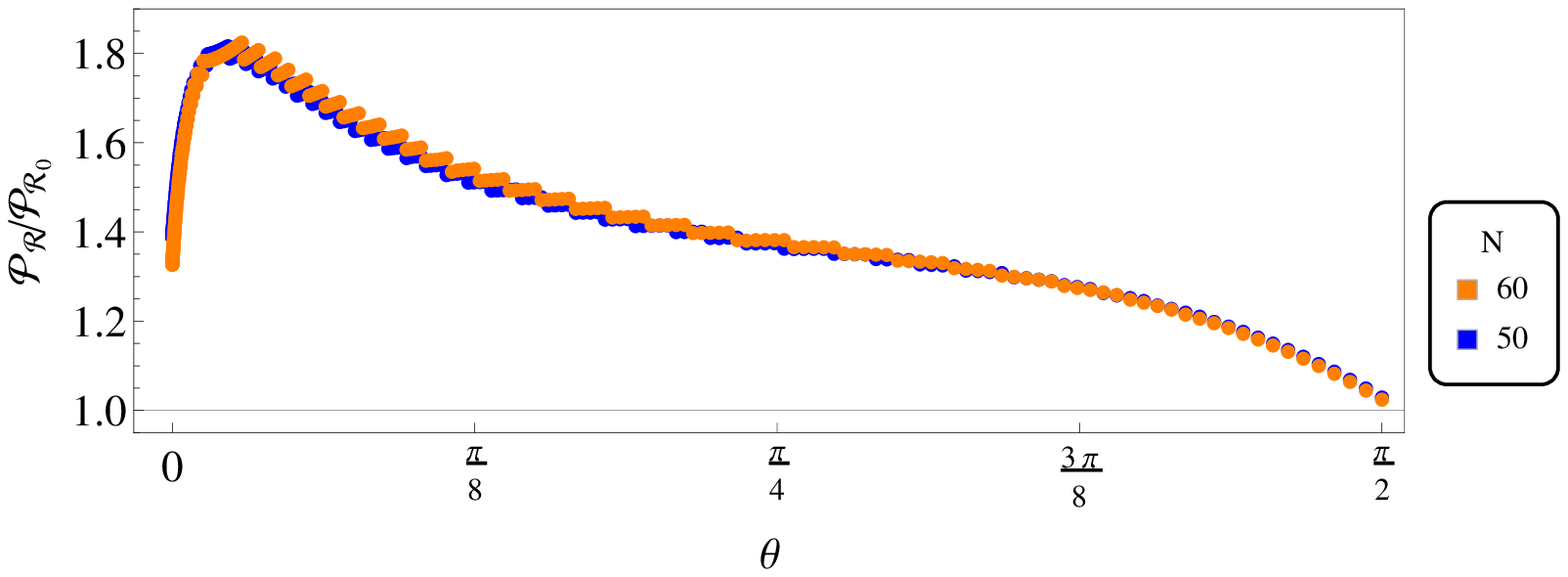}}
	\caption{\it The curvature power spectrum normalized to the single-field expression,
	$\mathcal{P}_{\mathcal{R}}$, for the mode that leaves the 
	horizon at the start of the last $N = 50$ (blue) or $60$ (orange) e-folds of inflation is shown as a 
	function of $\theta$  for $c=100$ (upper panel)
	and for $c= 0.1$ (lower panel).} \label{thps}
	\vspace{10pt}
\end{figure}

The enhancement of the power spectrum, $\mathcal{P}_{\mathcal{R}}$, is shown as a function of $\theta$ in Fig.~\ref{thps}
for $c = 100$ (upper panel) and $c = 0.1$ (lower panel). In the $c = 100$ case, the enhancement is generally $\lesssim 2$\%,
apart from a small enhancement to nearly 4\% for $\theta \sim \pi/2$, which is associated with the irregular behaviour of the
corresponding trajectory in the left panel of Fig.~\ref{thtraj}. As could be anticipated from the right panel in Fig.~\ref{thtraj},
the behaviour of $\mathcal{P}_{\mathcal{R}}$ for $c = 0.1$ is more regular, first rising from $\sim 1.4$ to $\sim 1.8$
and then decreasing towards negligible values as $\theta \to 0$.

The bottom part of Fig.~\ref{thps} corresponds to weak stabilization, $c=0.1$. 
In this case, the kinetic coupling between the real and imaginary parts of $T$ is still sufficiently strong to
cause the inflationary trajectory to deviate from the expected path along the valley of the inflationary potential, 
for small $\theta$. This is visible in the violet and blue trajectories in Fig.~\ref{thtraj}, which bend towards the real direction.
As a result, a peak in the power spectrum develops at $\theta \gtrsim 0$.
At $\theta=0$ the stabilization is purely quadratic (see Eq.~(\ref{ktheta})) and the power spectrum shows a milder enhancement.

In contrast, the power spectrum at $c=100$ shows a peak at $\theta\simeq \pi/2$,
which is a result of the weaker quartic stabilization that dominates near $\theta=\pi/2$. 
The scalar potential is very flat for $\rho\lesssim 3$, and the kinetic coupling drives the imaginary part $\alpha$
to values away from the minimum of the inflationary valley, As a result, two-field effects are important 
(see the red trajectory near $\theta=\pi/2$ in Fig.~\ref{thtraj}), and the power spectrum is enhanced. 
The large value of $c$ confines the motion of the fields to a narrow valley, 
and the enhancement is therefore not as dramatic as it is for small $c$. However,
at $\theta=\pi/2$ the kinetic coupling vanishes, and inflation is purely single-field.

For intermediate $c$, the power spectrum generally exhibits some combination of both features,
which become of comparable size for $c\simeq0.3$. For $c<0.3$ the largest enhancement is due to the 
weakness of the stabilization parameter, whereas for $c>0.3$ the peak is due to the weaker quartic stabilization
near the real axis. For $c = 0.3$, the power spectrum is enhanced by a factor between 1.1 and 1.2 over a 
broad range in $\theta$ between $\pi/40$ and $7\pi/16$.

\begin{figure}[!h]
\centering
	\scalebox{0.9}{\includegraphics{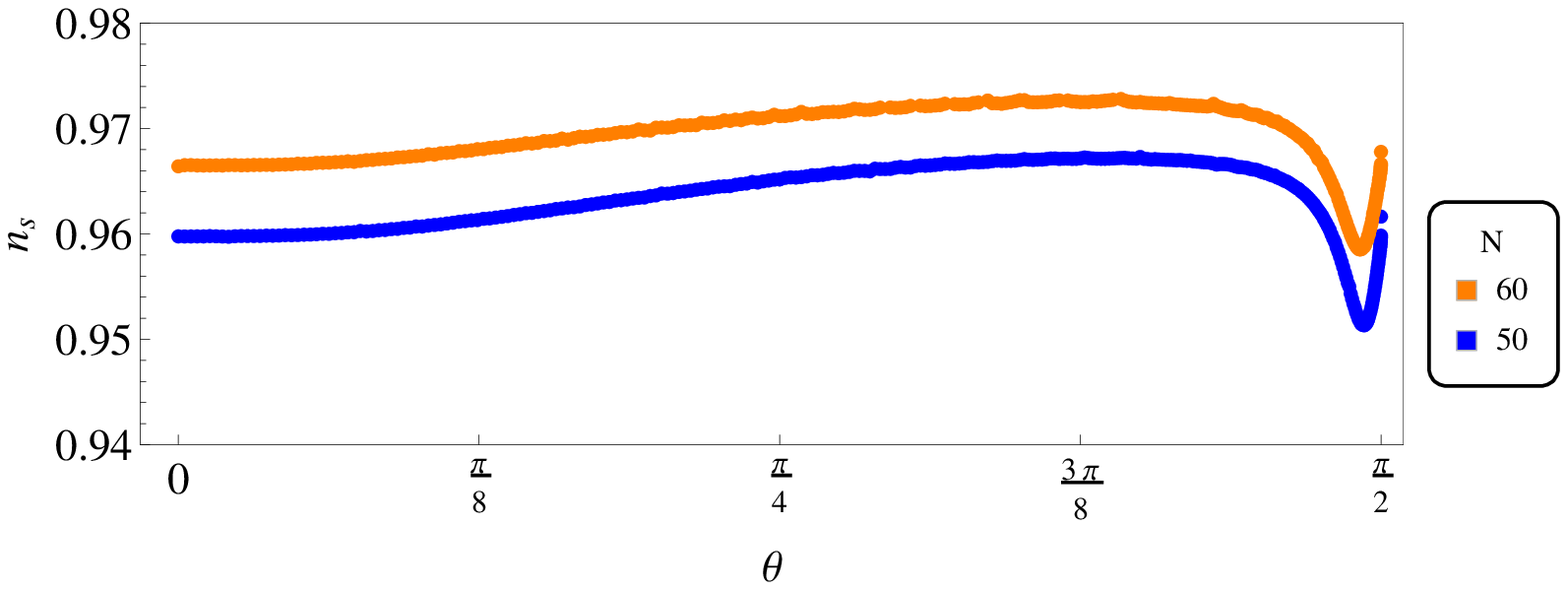}} \\
	\vspace{10pt}
	\scalebox{0.9}{\includegraphics{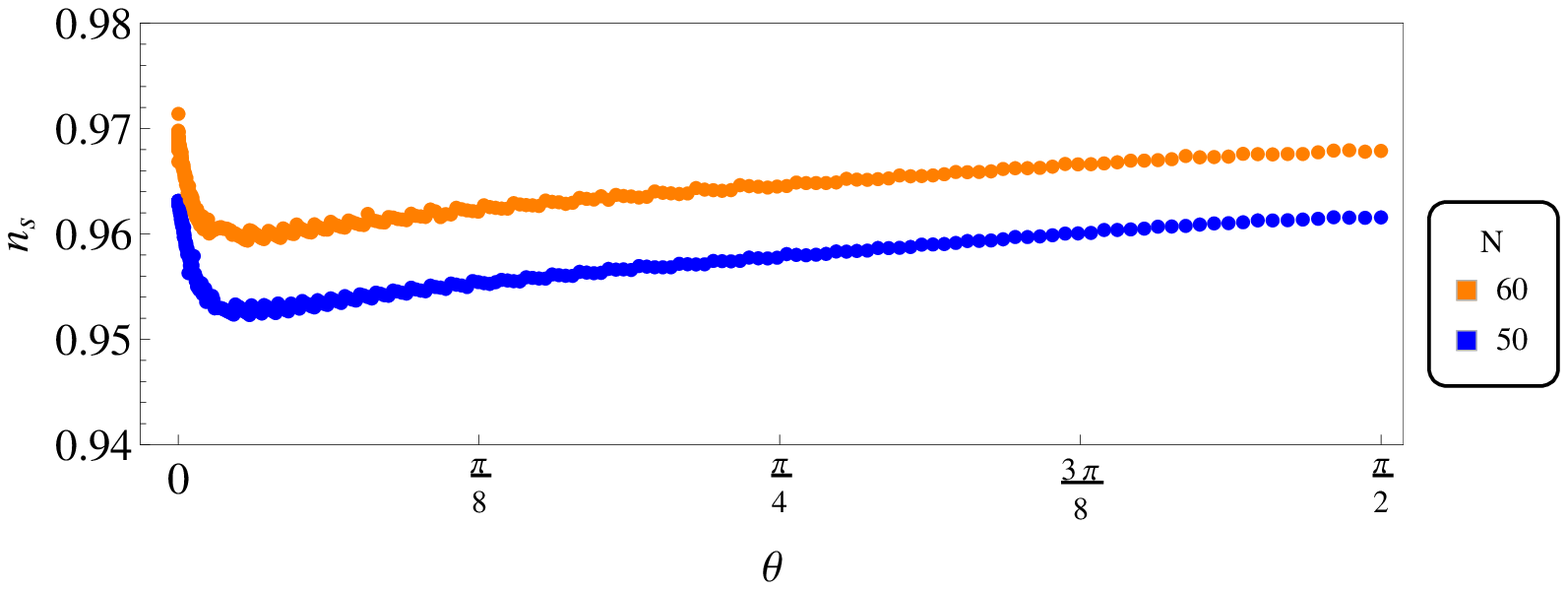}} 
	\caption{\it Dependences of the spectral index $n_s$ on the angle $\theta$ at $N = 50$ (blue lines)
	and $60$ e-folds (orange lines) for $c=100$ (upper panel) and $c = 0.1$ (lower panel).} \label{thns1}
\end{figure}

The corresponding values of $n_s$ at $N = 50$ (blue) and 60 (orange) e-folds are shown in Fig.~\ref{thns1}
for $c = 100$ (upper panel) and $c = 0.1$ (lower panel). When $c = 100$, we find that $0.95 \lesssim n_s \lesssim 0.97$
for both the displayed values of $N$ at all values of $\theta$. The only notable feature is a downward
glitch at $\theta \sim \pi/2$, which is associated with the corresponding excursion in the inflaton
trajectory seen in the left panel of Fig.~\ref{thtraj}. In contrast, when $c = 0.1$
$n_s$ initially decreases for small $\theta$ and then increases monotonically as $\theta \to \pi/2$, with
values in the acceptable range $0.95 \lesssim n_s \lesssim 0.97$ for both $N = 50$ and $N = 60$.

\begin{figure}[!h]
\centering
	\scalebox{0.9}{\includegraphics{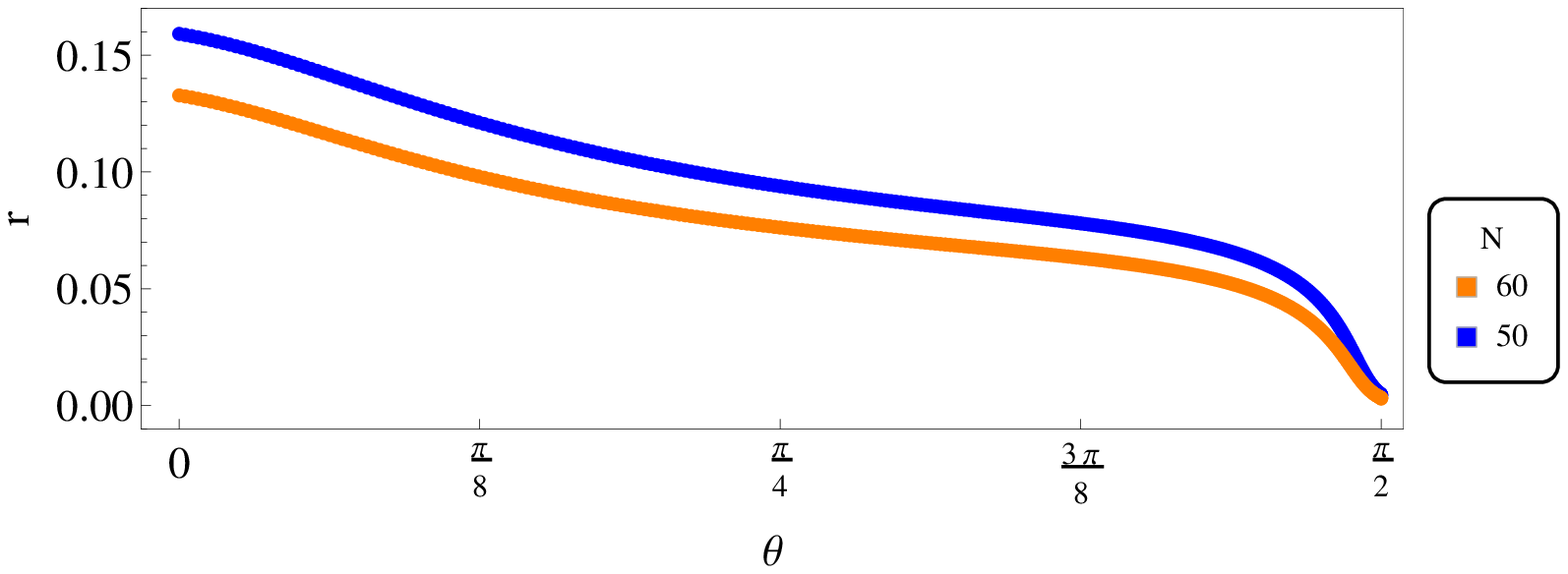}} \\
	\vspace{10pt}
	\scalebox{0.9}{\includegraphics{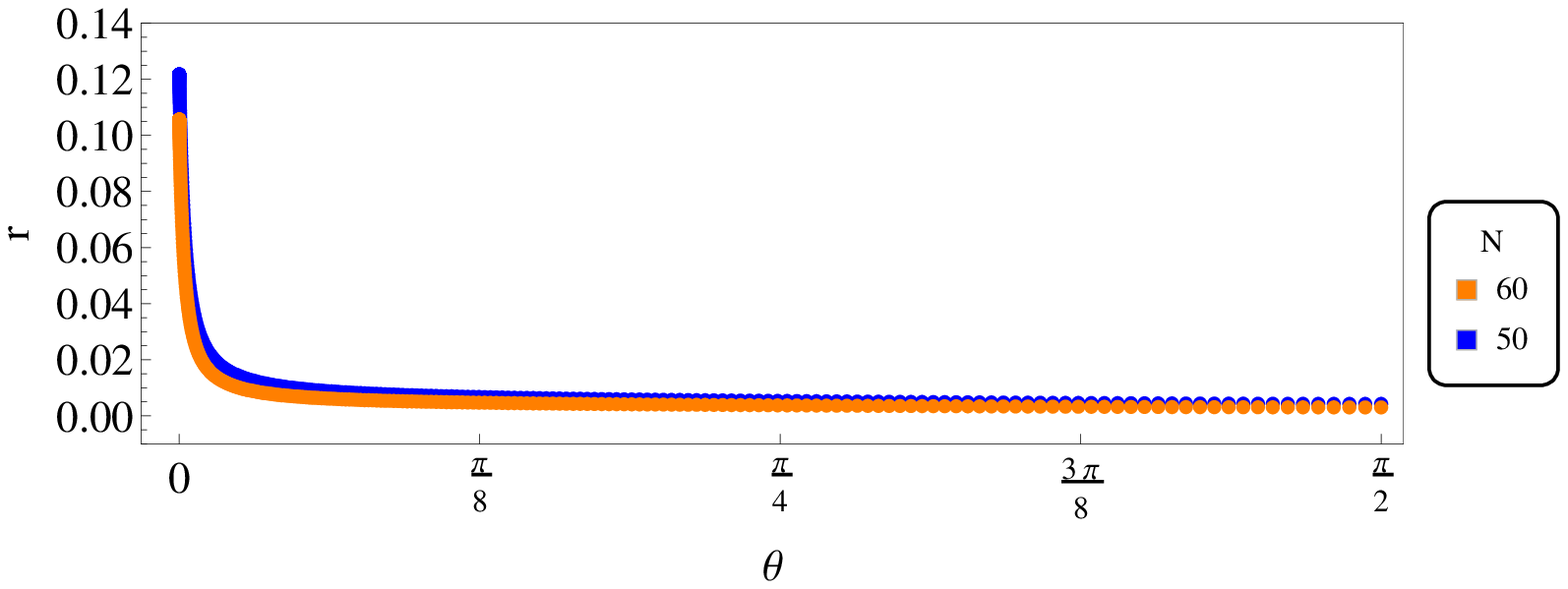}} 
	\caption{\it Dependences of the tensor-to-scalar ratio $r$ on the angle $\theta$ at $N = 50$ (blue lines)
	and $60$ e-folds (orange lines) for $c=100$ (upper panel) and $c = 0.1$ (lower panel).} \label{thr1}
\end{figure}

The corresponding values of $r$ at $N = 50$ (blue) and 60 (orange) e-folds are shown in Fig.~\ref{thr1}
for $c = 100$ (upper panel) and $c = 0.1$ (lower panel). In the large-$c$ case, we see a smooth
interpolation between BICEP2- and Planck-friendly values of $r$ as $\theta$ increases from $0 \to \pi/2$.
In the small-$c$ case, the value of $r$ is small except for very small $\theta$, and is in fact below
0.03 for most values of $\theta$.

\begin{figure}[!h]
\centering
	\scalebox{0.49}{\includegraphics{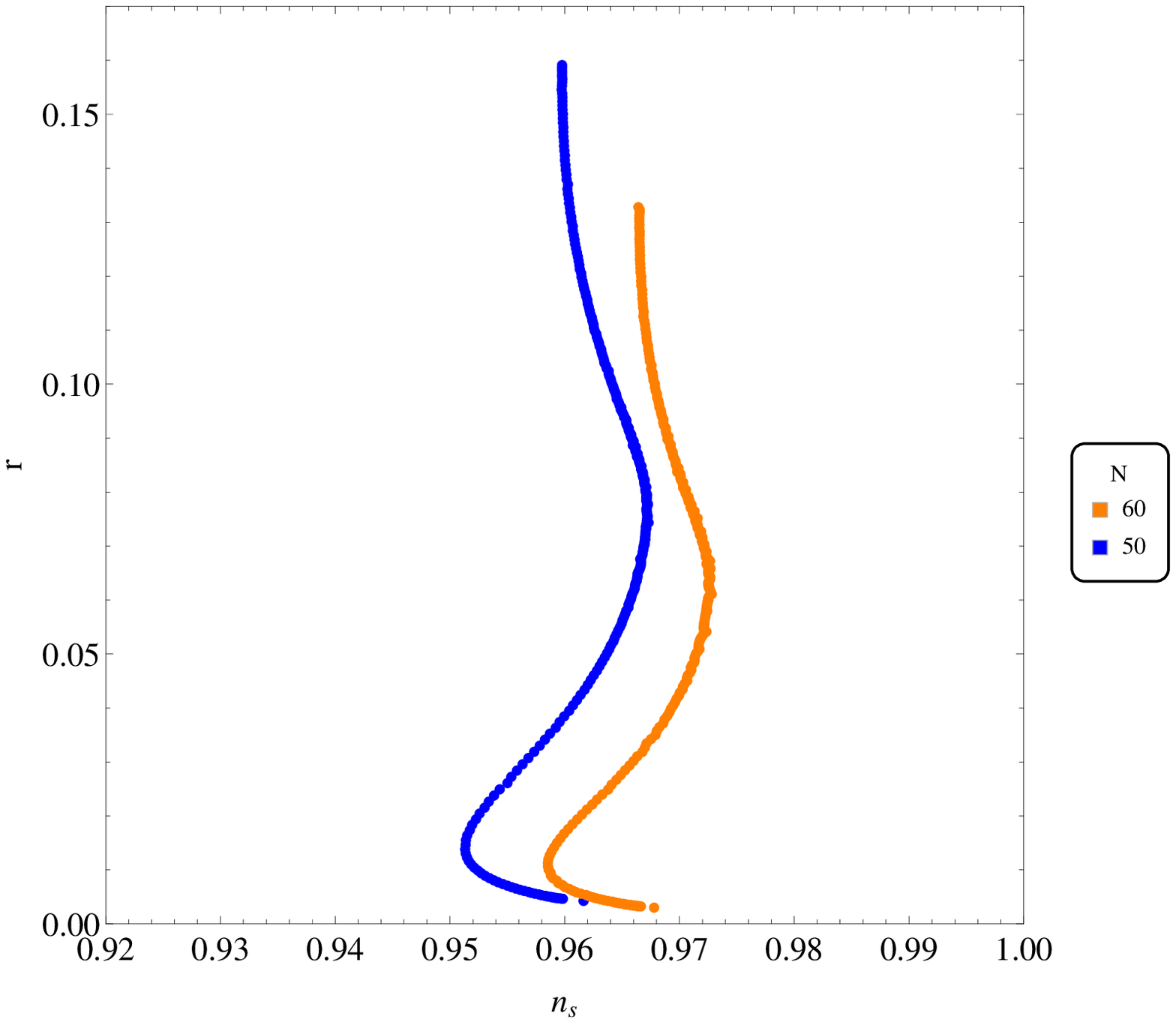}} 
	\scalebox{0.49}{\includegraphics{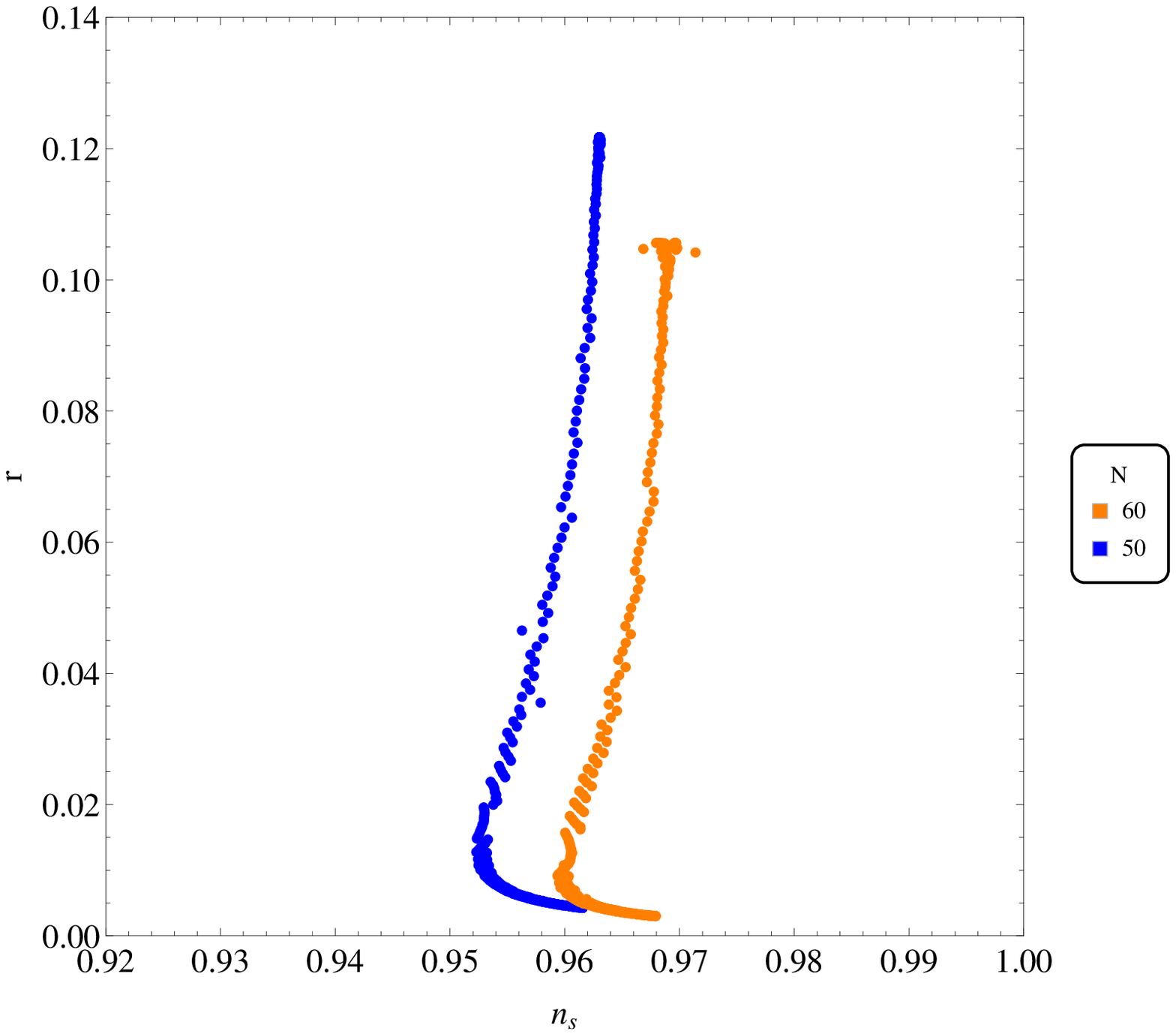}}
	\caption{\it Parametric curve $(n_s(\theta),r(\theta))$ for $c=100$ (left panel) and $c = 0.1$
	(right panel), both for $N=50$ (blue) and $60$ (orange) e-folds.} \label{thnsr2}
\end{figure}

The correlations in the behaviours of $n_s$ and $r$ are seen in Fig.~\ref{thnsr2},
which displays the parametric curves $(n_s(\theta),r(\theta))$ for $N = 50$ (blue lines)
and $60$ e-folds (orange lines) for $c=100$ (left panel) and $c = 0.1$ (right panel).
It is clear from the left panel that an interpolation between BICEP2- and Planck-friendly values of
$(n_s, r)$ is possible with $c\gg1$ as argued in~\cite{EGNO2}, even when two-field effects are
taken into account. On the other hand, as seen in the right panel, when $c = 0.1$ only small,
Planck-friendly values of $r$ are found.

\begin{figure}[!h]
\centering
	\scalebox{0.49}{\includegraphics{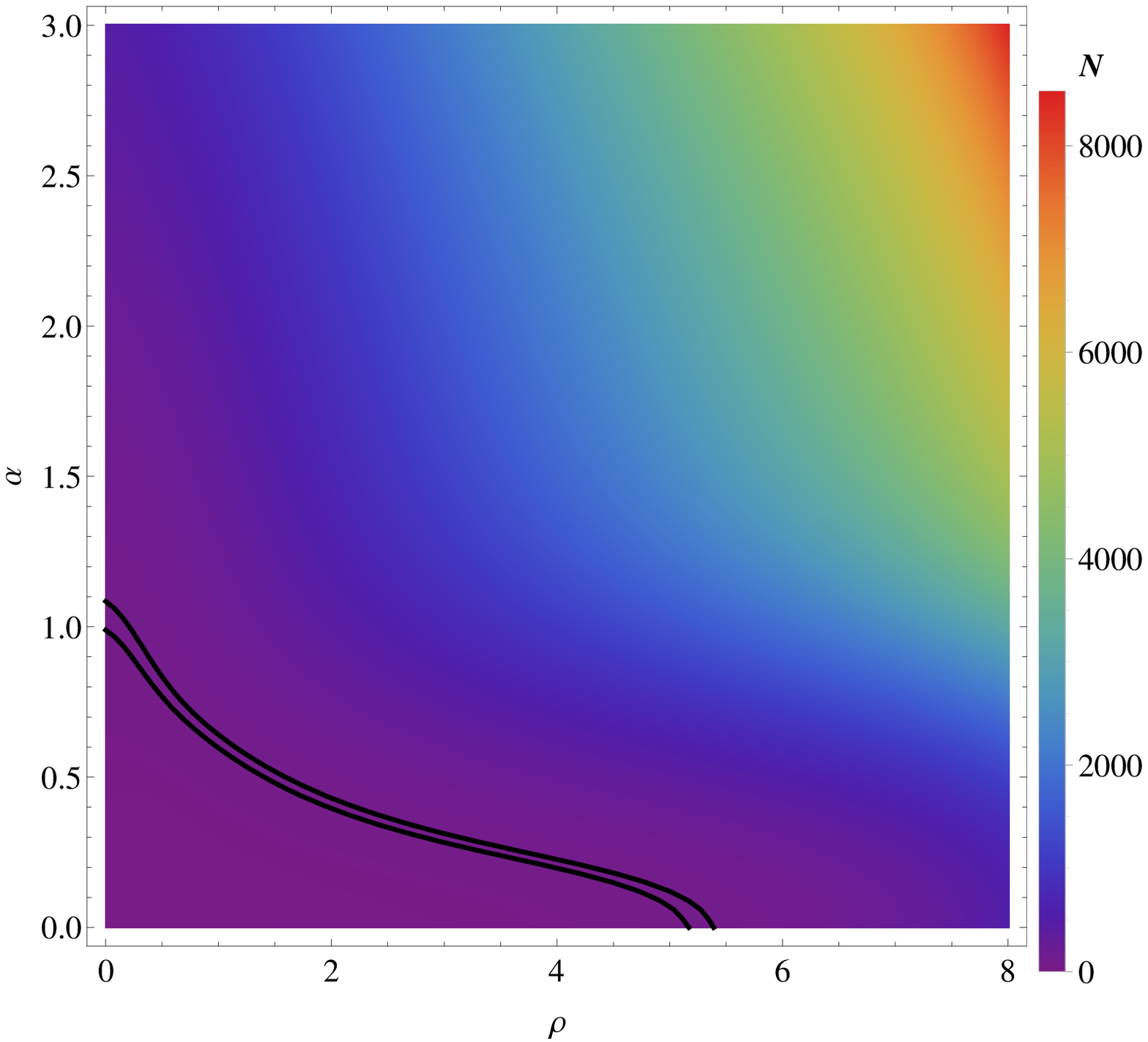}}
		\scalebox{0.49}{\includegraphics{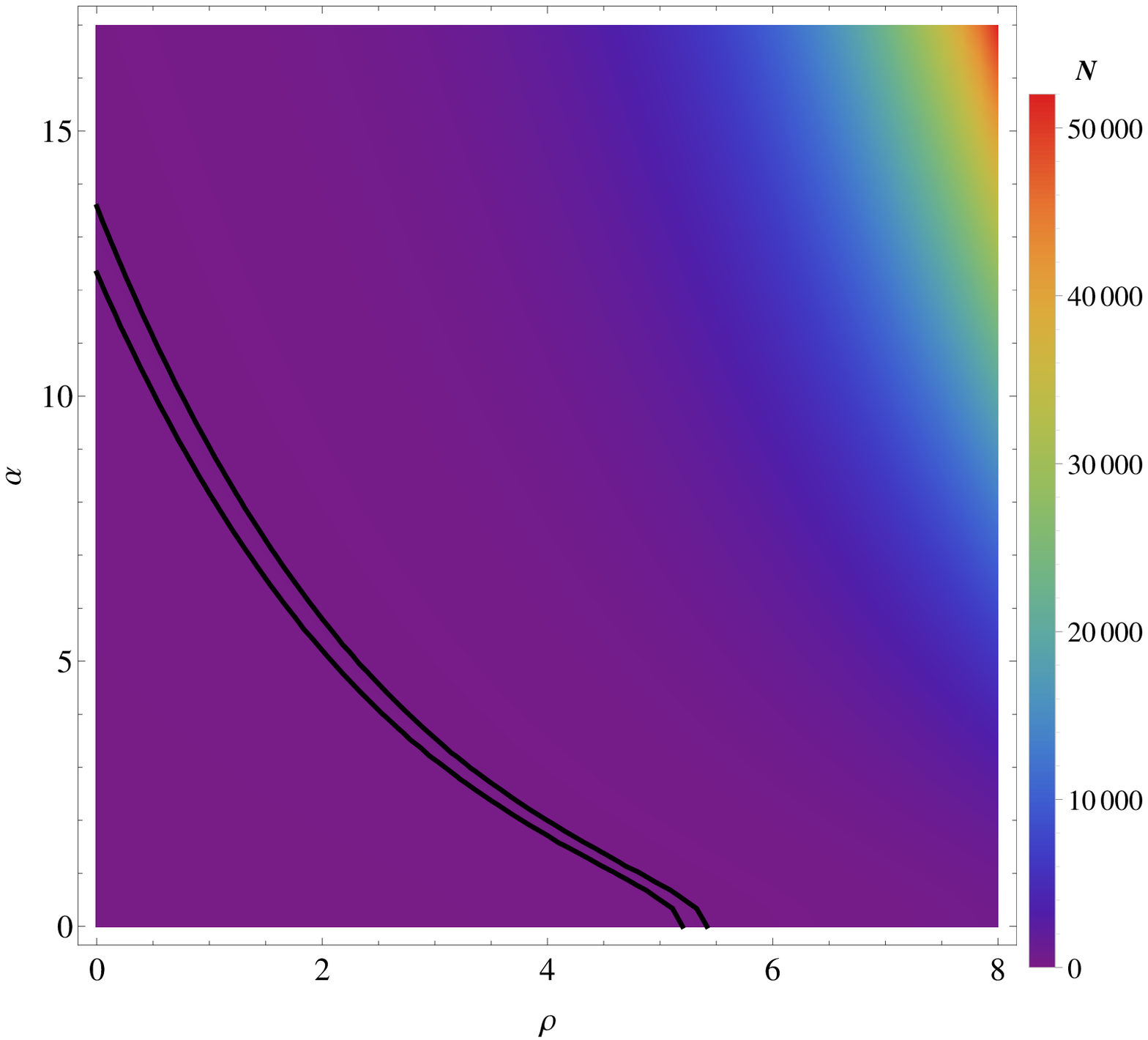}} 
	\caption{\it The total number of e-folds $N_{\rm tot}$ as a function of the fields in the $(\rho,\alpha)$ plane, for $c=100$
	(left panel) and $c = 0.1$ (right panel).
	The solid black lines correspond to $N=50$ (lower) and $N=60$ (upper).} \label{Nplane}
	\vspace{40pt}
\end{figure}

All the above results are for a total number $N_{\rm tot}=70$ of e-folds, and
we now explore the possible modifications of the above results for larger values of $N_{\rm tot}$.
Fig.~\ref{Nplane} shows the total number of e-folds $N_{\rm tot}$ as a function of the initial conditions in the $(\rho,\alpha)$ plane
for $c = 100$ (left panel) and $c = 0.1$ (right panel), with solid lines corresponding to $N=50$ (lower)
and $N=60$ (upper). Each point in the plane corresponds to the particular form of the K\"ahler potential (and therefore of the scalar potential) for which the stabilizing term vanishes initially, 
\beq
\tan\theta = \frac{3}{2\alpha_0^2}\left(e^{-\sqrt{\frac{2}{3}}\rho_0}-1\right),
\eeq
i.e., the initial condition is chosen at the bottom of the inflationary valley. We see that when $c = 0.1$ the 
variation of $N_{\rm tot}$ in the displayed part
of the $(\rho,\alpha)$ plane is much greater than for $c=100$, and much larger values of $N_{\rm tot}$
may be attained, though one should bear in mind that the canonical field along the imaginary axis is 
related to $\alpha$ in a $c$-dependent way as shown in Eq. (\ref{canon}).

\begin{figure}[!h]
\centering
	\hspace{-15pt}
	\scalebox{0.49}{\includegraphics{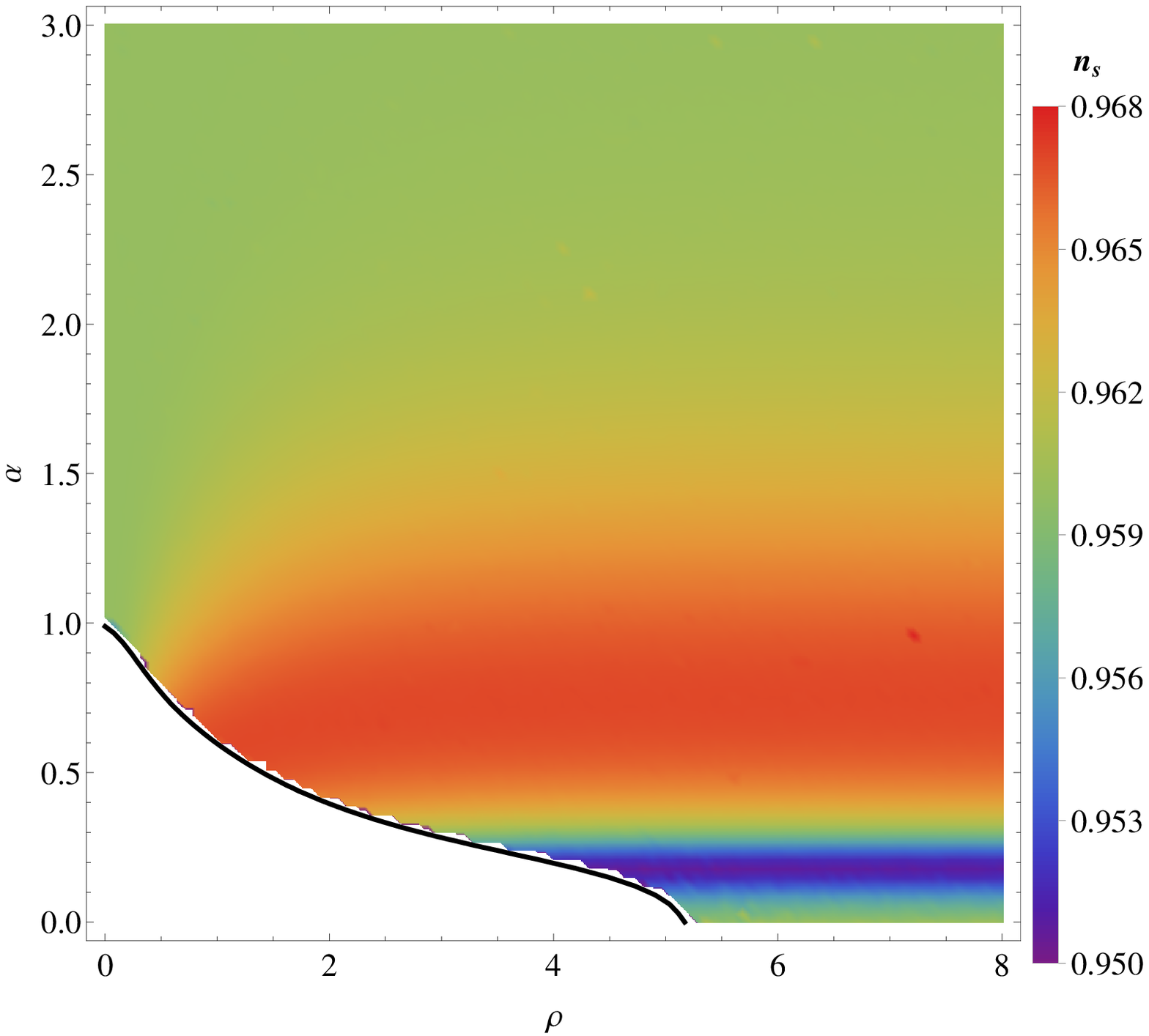}}
	\scalebox{0.49}{\includegraphics{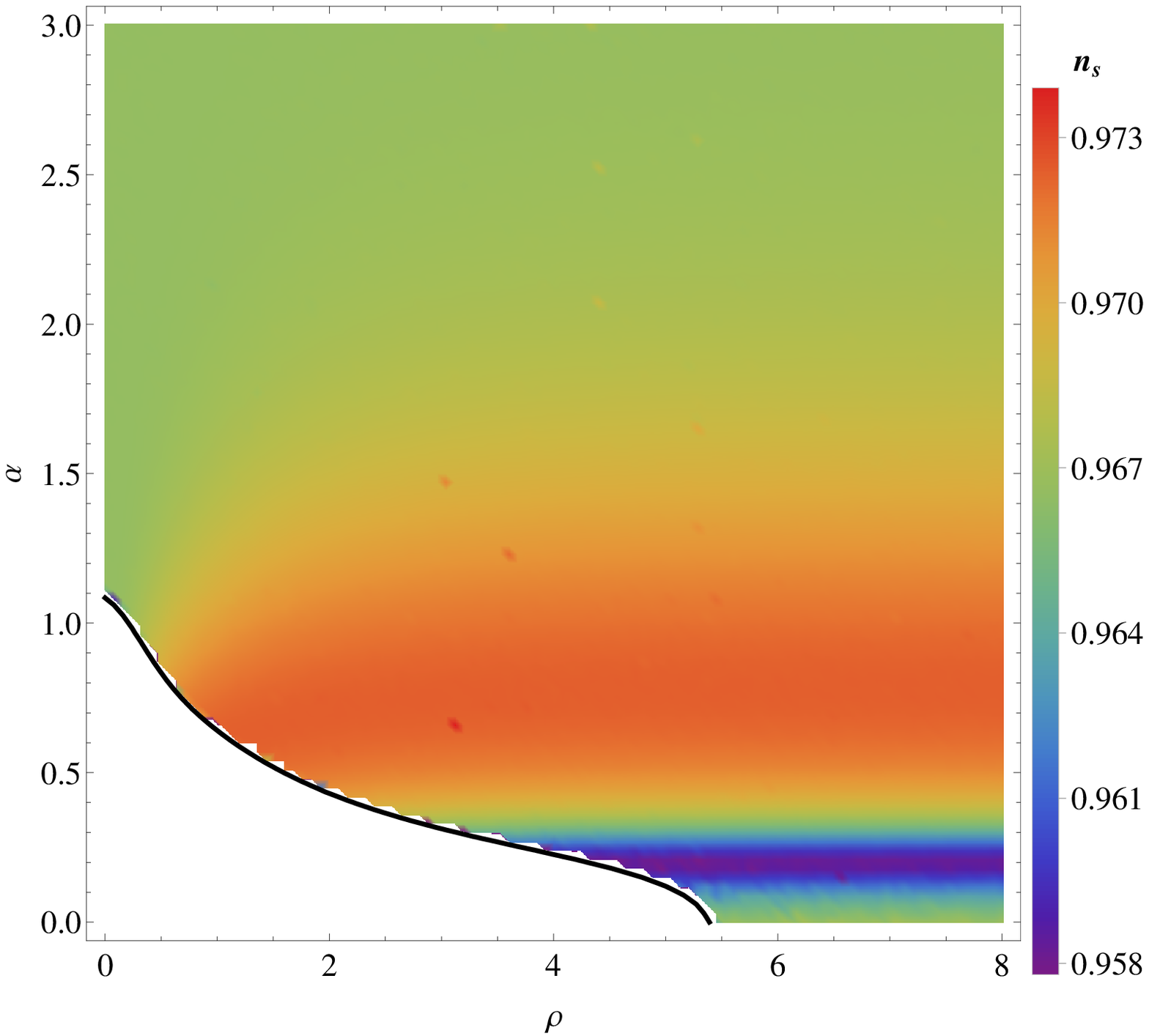}} \\
	\scalebox{0.49}{\includegraphics{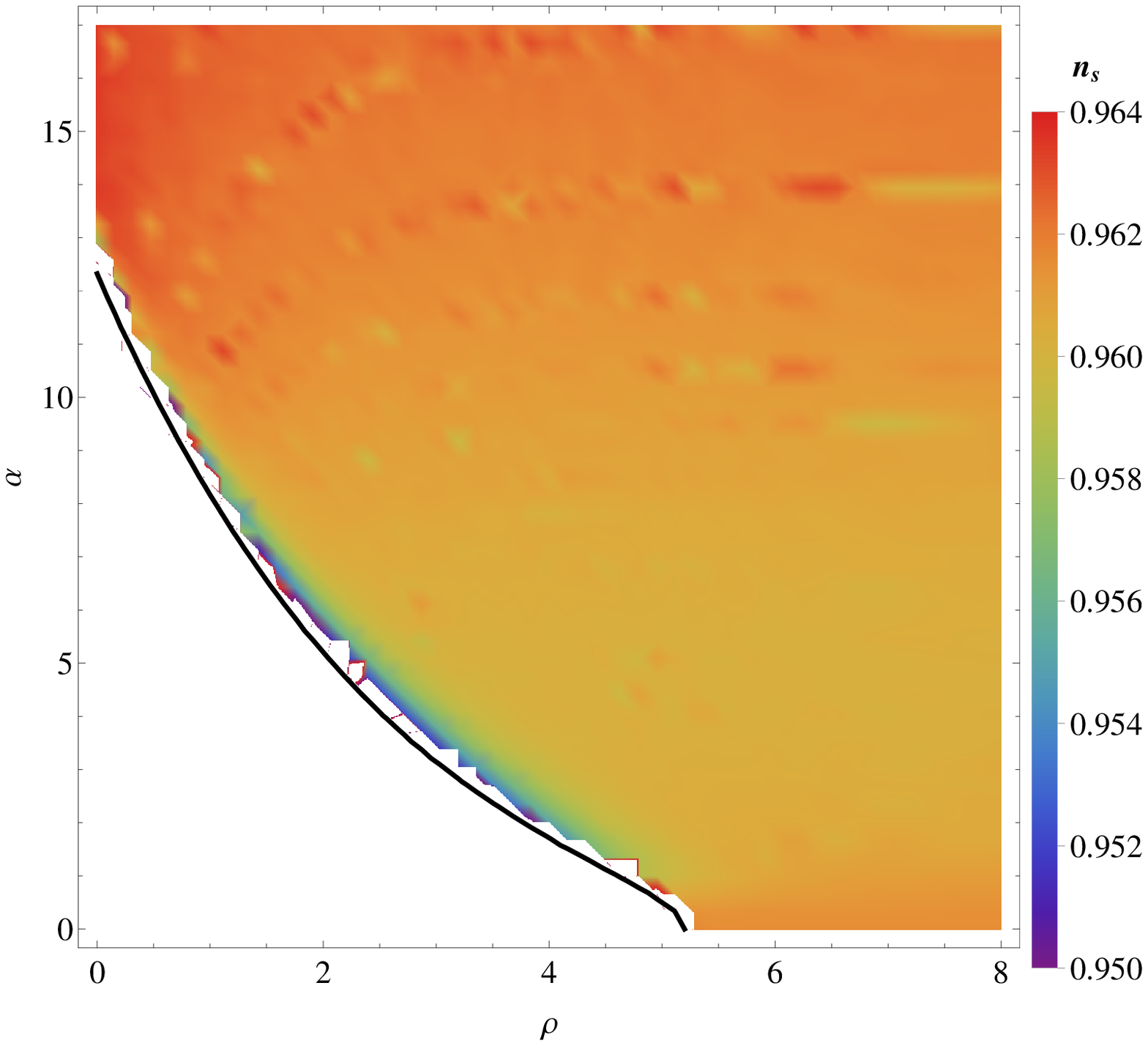}}
	\scalebox{0.49}{\includegraphics{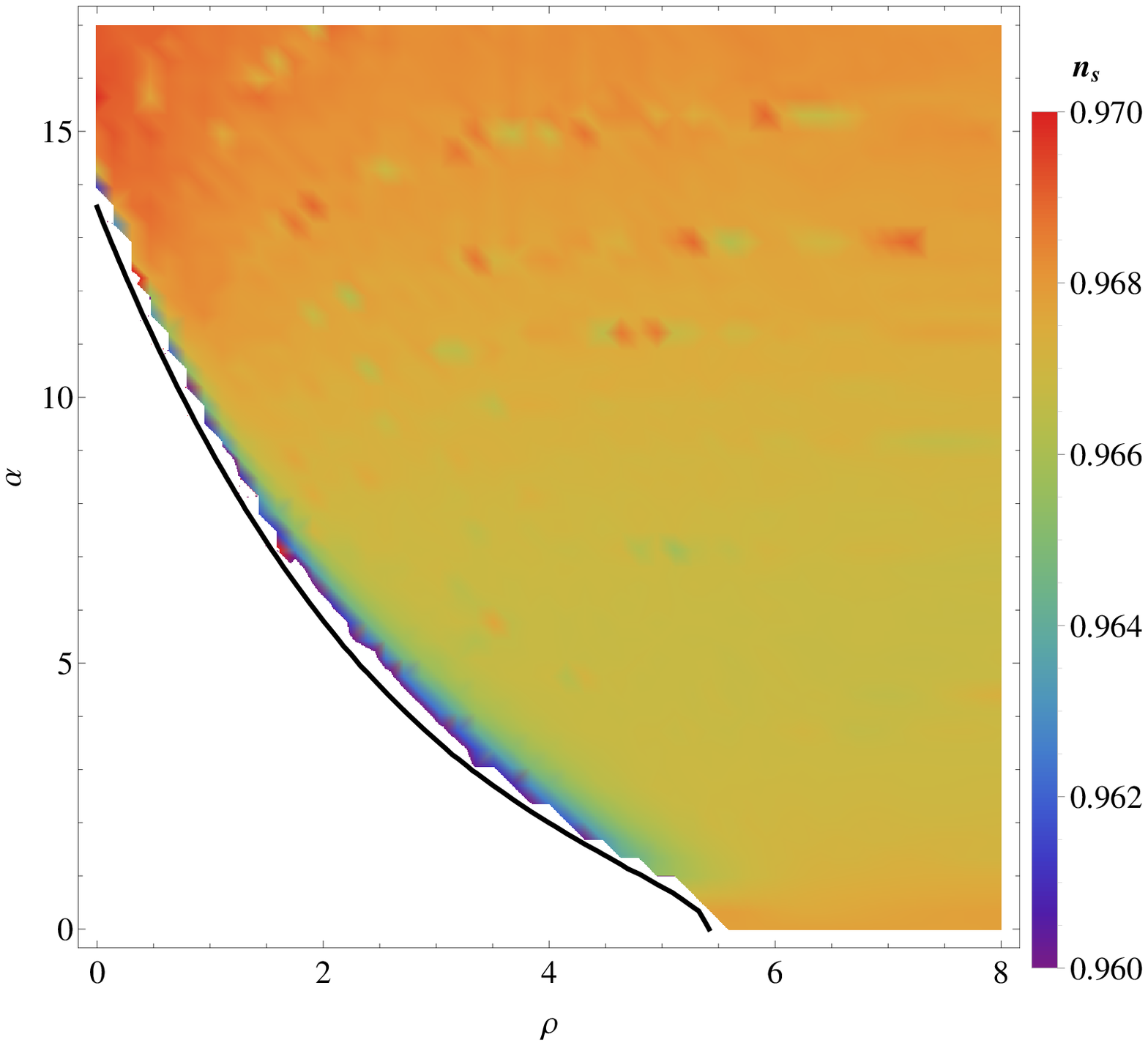}}
	\caption{\it Scalar spectral index $n_s$ from the two-field analysis in the $(\rho,\alpha)$ plane,
	evaluated at $N = 50$ (left panels) and $60$ (right panels) e-folds, with $c = 100$ (upper panels)
	and $c = 0.1$ (lower panels).
	The solid curves are the boundaries for the corresponding values of $N$.} \label{nsplane}
\end{figure}

Fig.~\ref{nsplane} displays general results for $n_s$ in the $(\rho,\alpha)$ plane. The upper (lower)
panels are for $c = 100$ ($c = 0.1$), and the left (right) panels are for $N = 50 (60)$ e-folds. In the
case $N = 50$ and $c = 100$, we see that there is a large region of the $(\rho,\alpha)$ plane with $\alpha \gtrsim 1.5$
where $n_s \simeq 0.960$. Larger values or $n_s \lesssim 0.970$ are found in a band around $\alpha \sim 1.0$,
and smaller values $n_s \gtrsim 0.950$ are found for $\alpha < 0.5$. Acceptable values of $n_s$ are found
throughout the $(\rho,\alpha)$ plane for $N = 50, c = 100$. The general
behaviour in the $N = 60, c = 100$ case is very similar, but the values of $n_s$ are larger by $\sim 0.008$
corresponding approximately to one current standard deviation, and there is a band with $\alpha \sim 1.0$
where $n_s \gtrsim 0.970$. In the case $N = 50, c = 0.1$, there is relatively little variation in $n_s$, with values
generally $\simeq 0.960$. There is also relatively little variation in $n_s$ across the $(\rho,\alpha)$ plane for
$N = 60, c = 0.1$, but the values are again generally larger by $\sim 0.008$.

\begin{figure}[!h]
\centering
	\hspace{-15pt}
	\scalebox{0.49}{\includegraphics{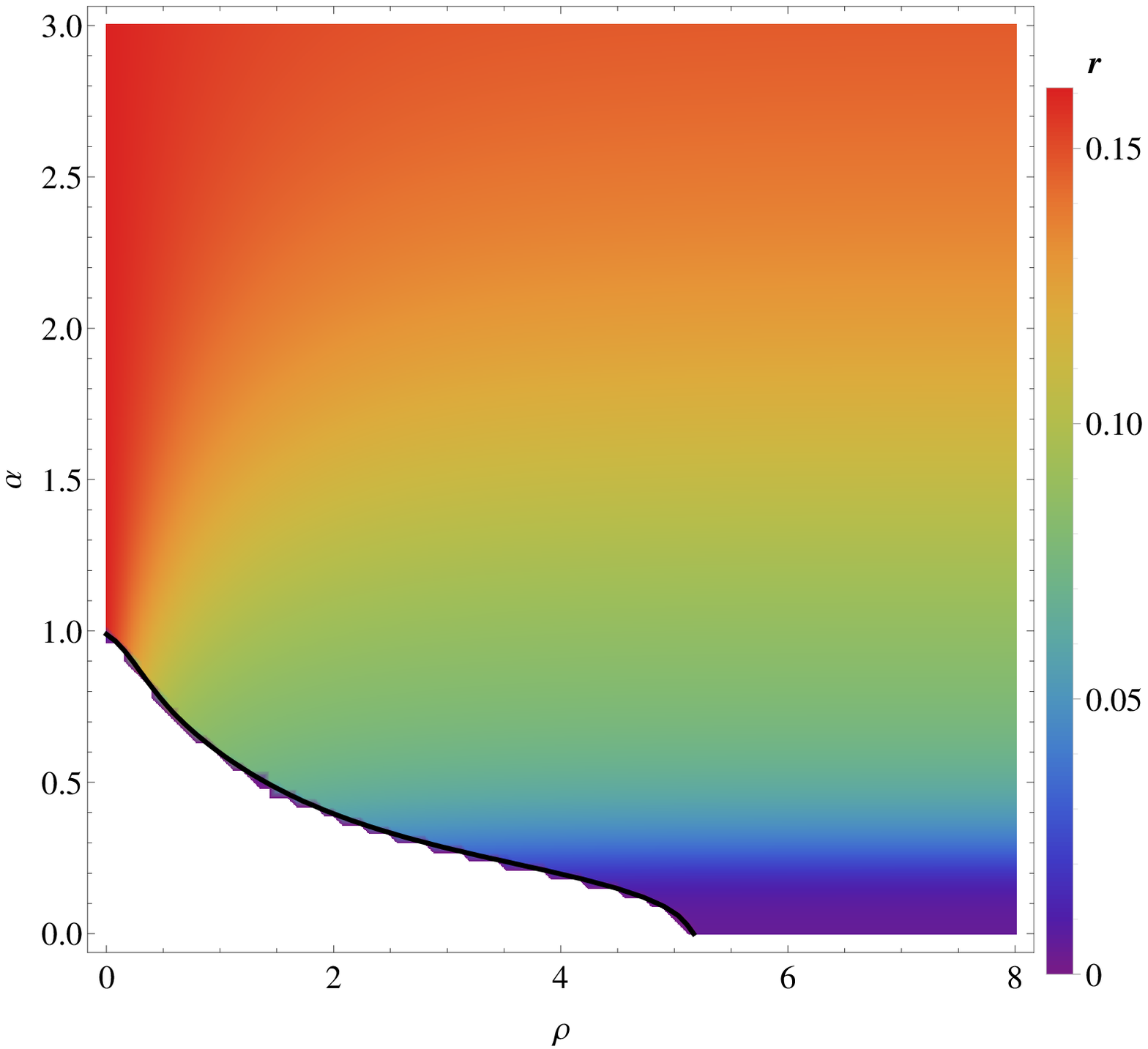}}
	\scalebox{0.49}{\includegraphics{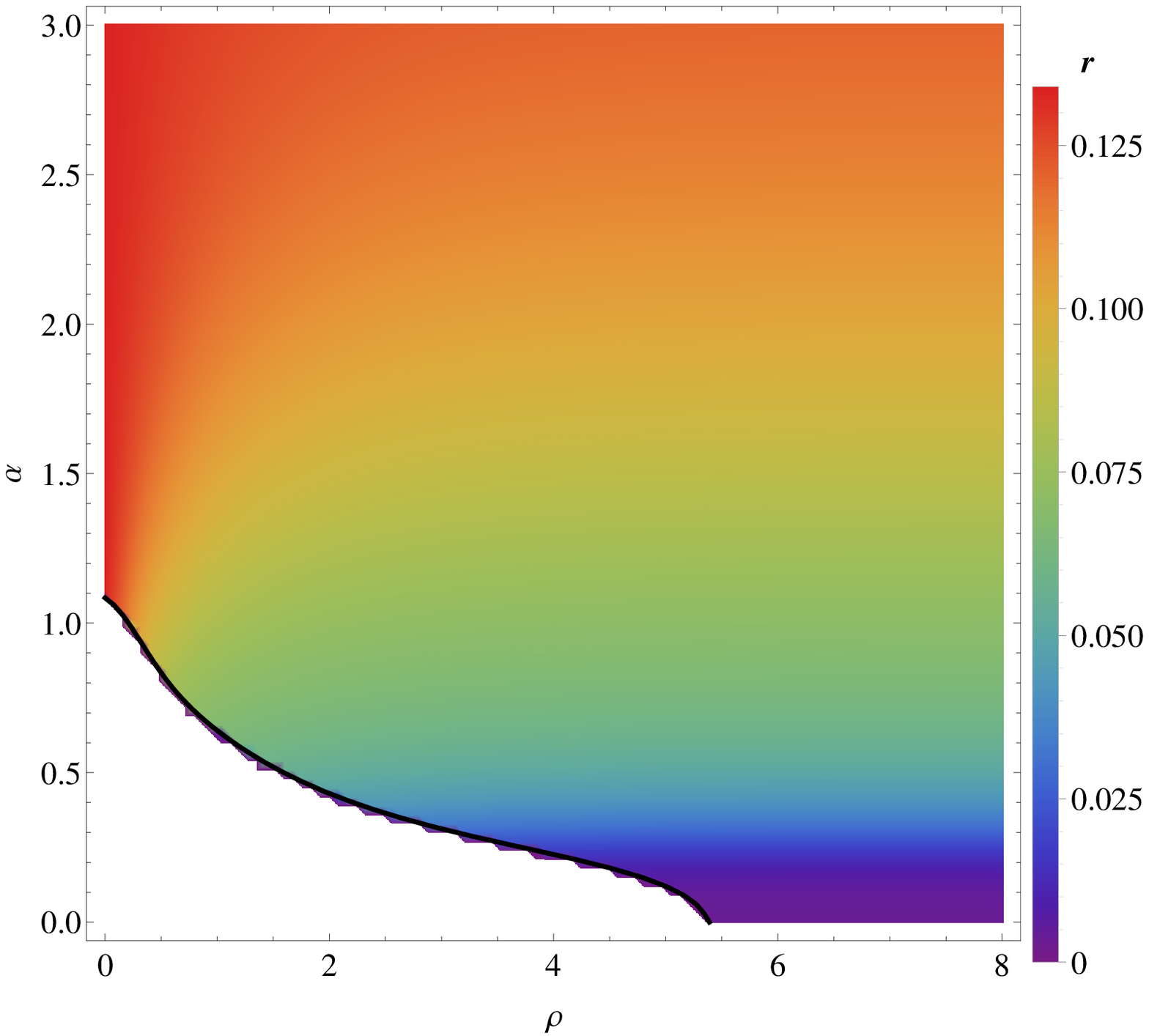}} \\
	\scalebox{0.49}{\includegraphics{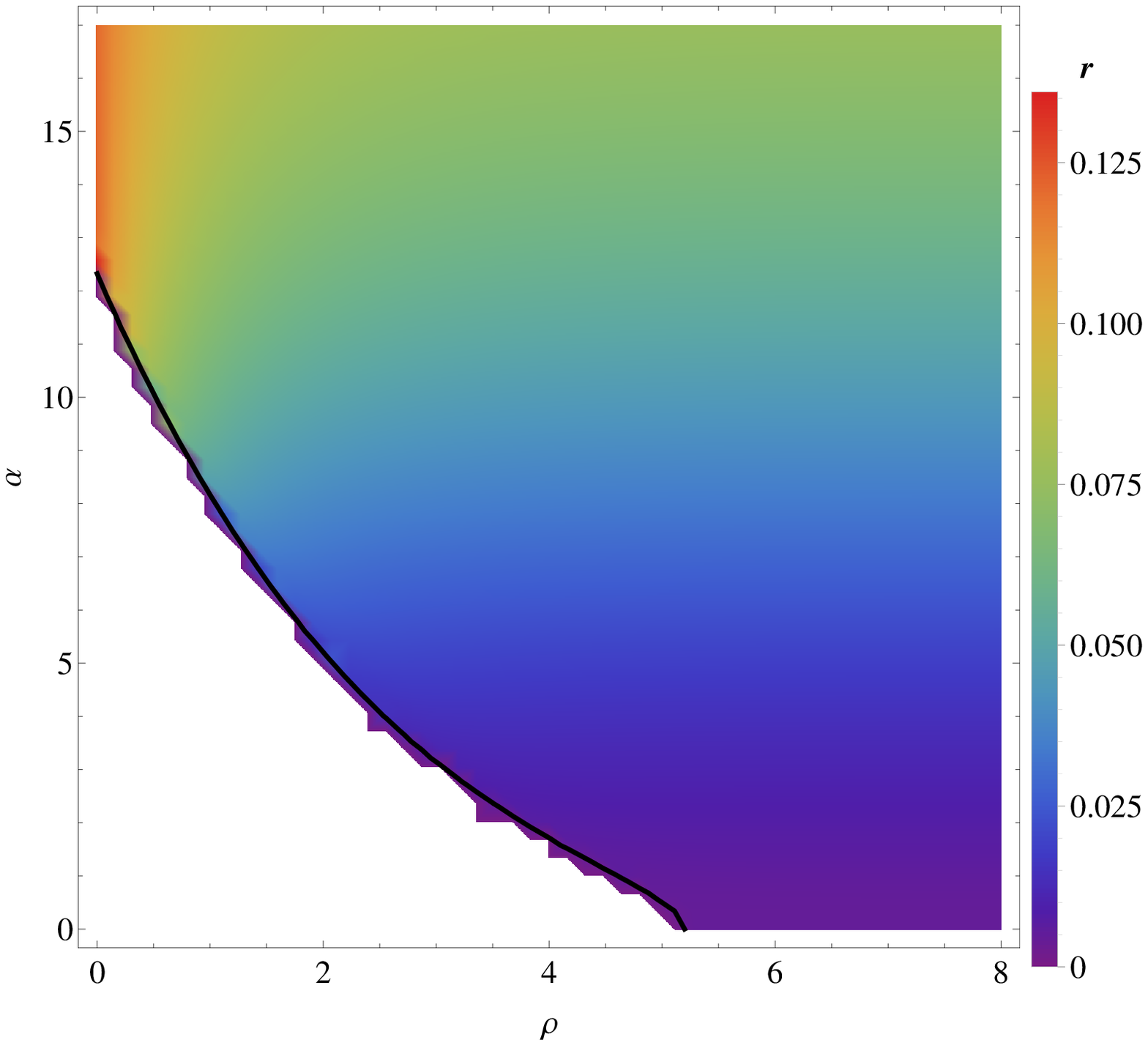}}
	\scalebox{0.49}{\includegraphics{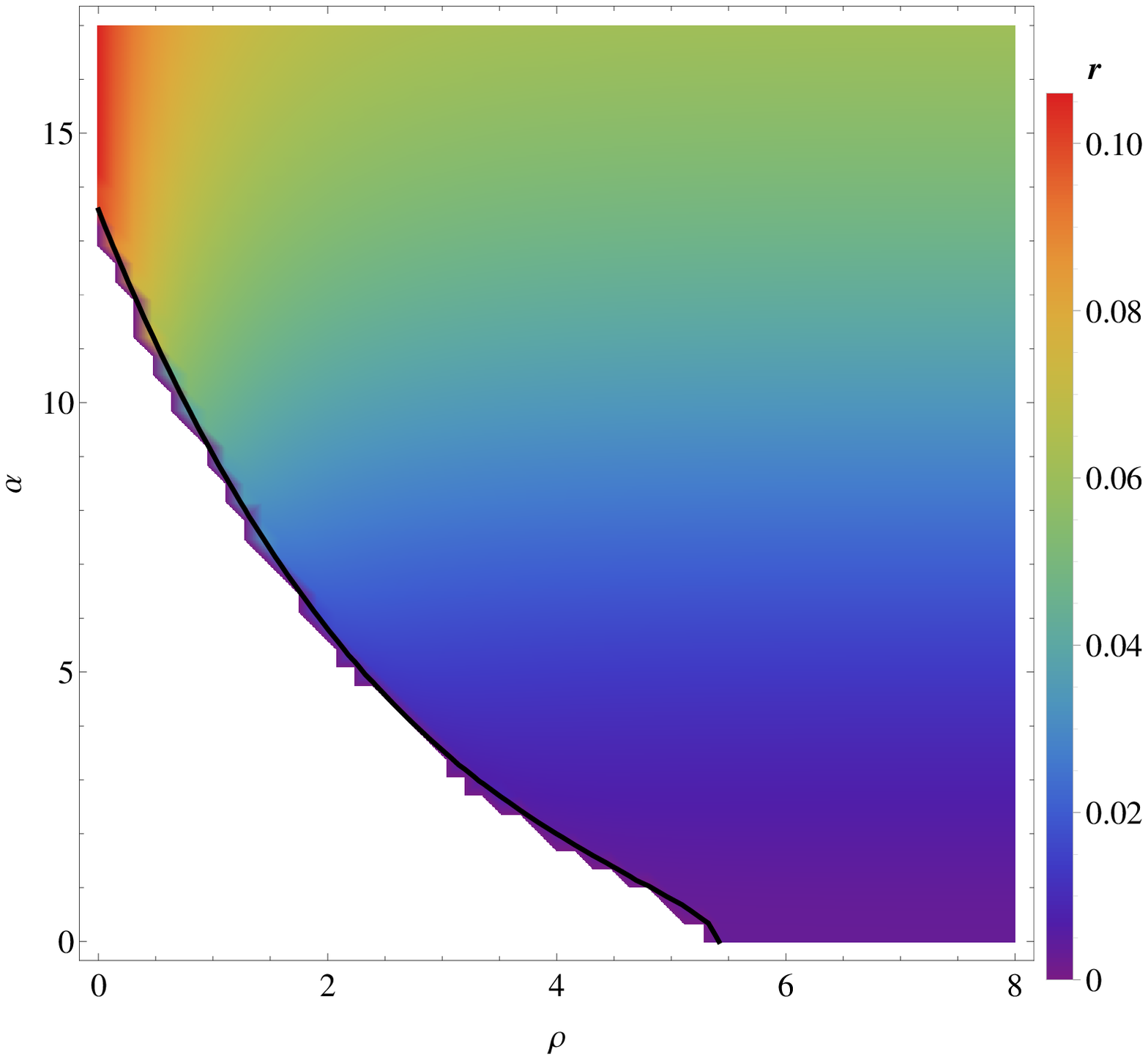}}
	\caption{\it Tensor-to-scalar ratio $r$ from the two-field analysis in the $(\rho,\alpha)$ plane,
	evaluated at $N = 50$ (left panels) and $60$ (right panels) e-folds, with $c = 100$ (upper panels)
	and $c = 0.1$ (lower panels).
	The solid curves are the boundaries for the corresponding values of $N$.} \label{rplane}
\end{figure}

Fig.~\ref{rplane} shows corresponding results for $r$ in the $(\rho,\alpha)$ plane.
In the $c = 100$ cases (upper panels), we see that BICEP2-friendly values of $r \gtrsim 0.10$
are found for $\alpha \gtrsim 1.7$ for all values of $\rho$, extending down to $\alpha \sim 1$ for
$\rho \lesssim 1$. We also see that $r \lesssim 0.03$ for $\alpha < 0.3$.
In the case $N = 50, c = 0.1$, we see that values of $r$ interpolate smoothly between the BICEP2-friendly
values found at small $\theta$ and the Planck-friendly values found when $\theta \sim \pi/2$. The values of $r$
for $N = 60, c = 0.1$ are slightly lower, but also interpolate smoothly between BICEP2- and Planck-friendly values.

\section{Non-Gaussianity}

\begin{figure}[!h]
\centering
	\hspace{-15pt}
	\scalebox{0.49}{\includegraphics{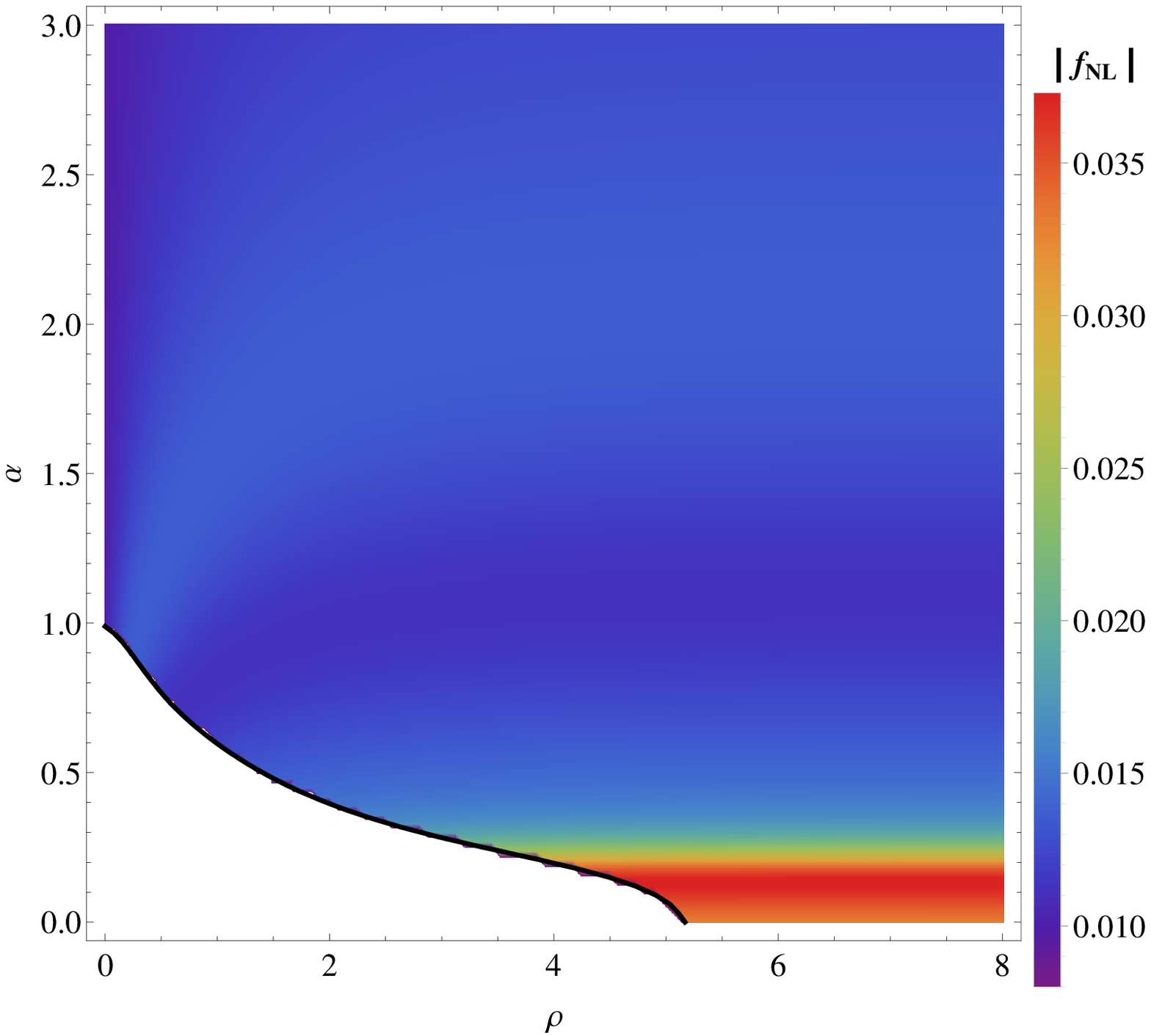}}
	\scalebox{0.49}{\includegraphics{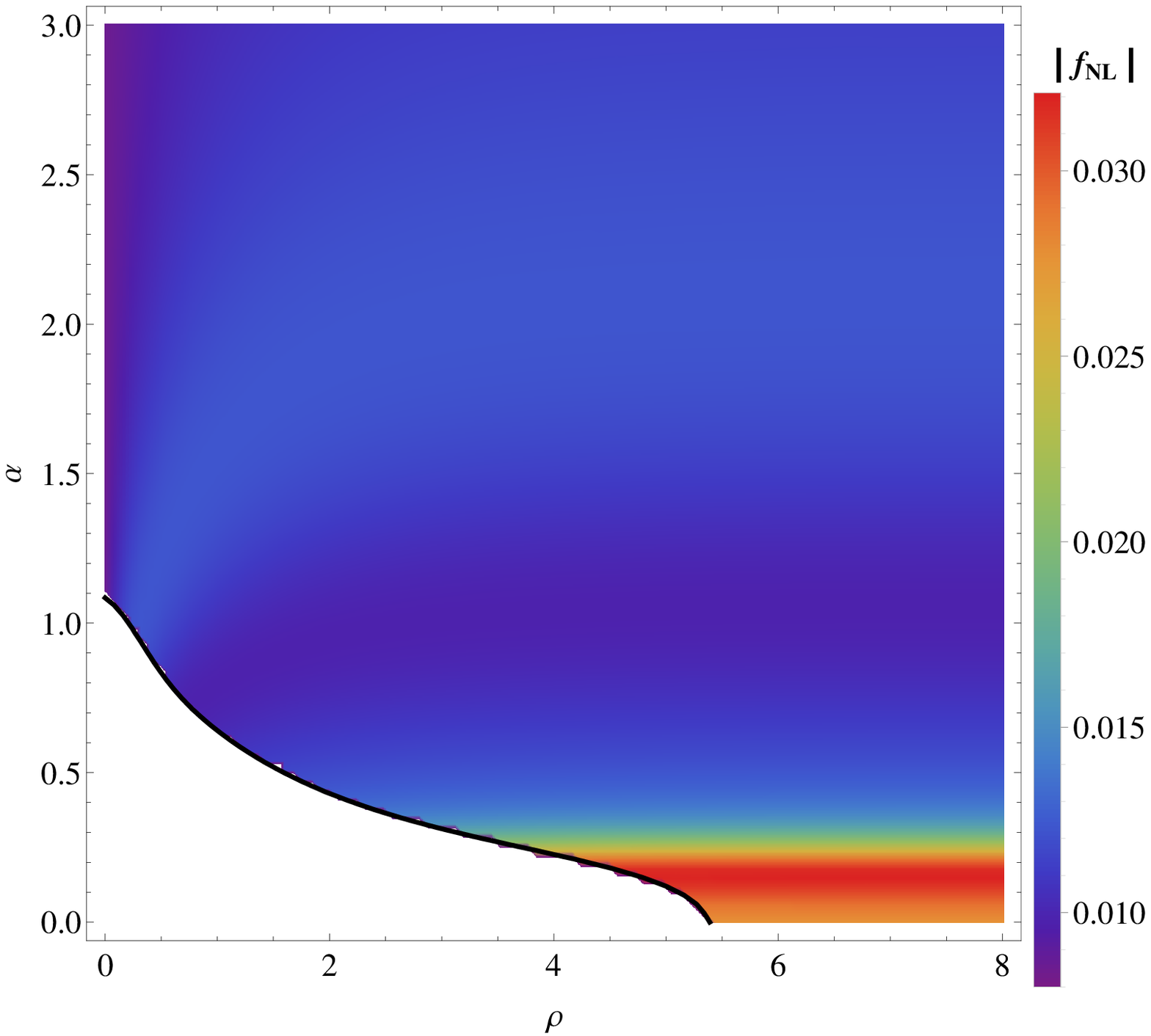}} \\
	\scalebox{0.49}{\includegraphics{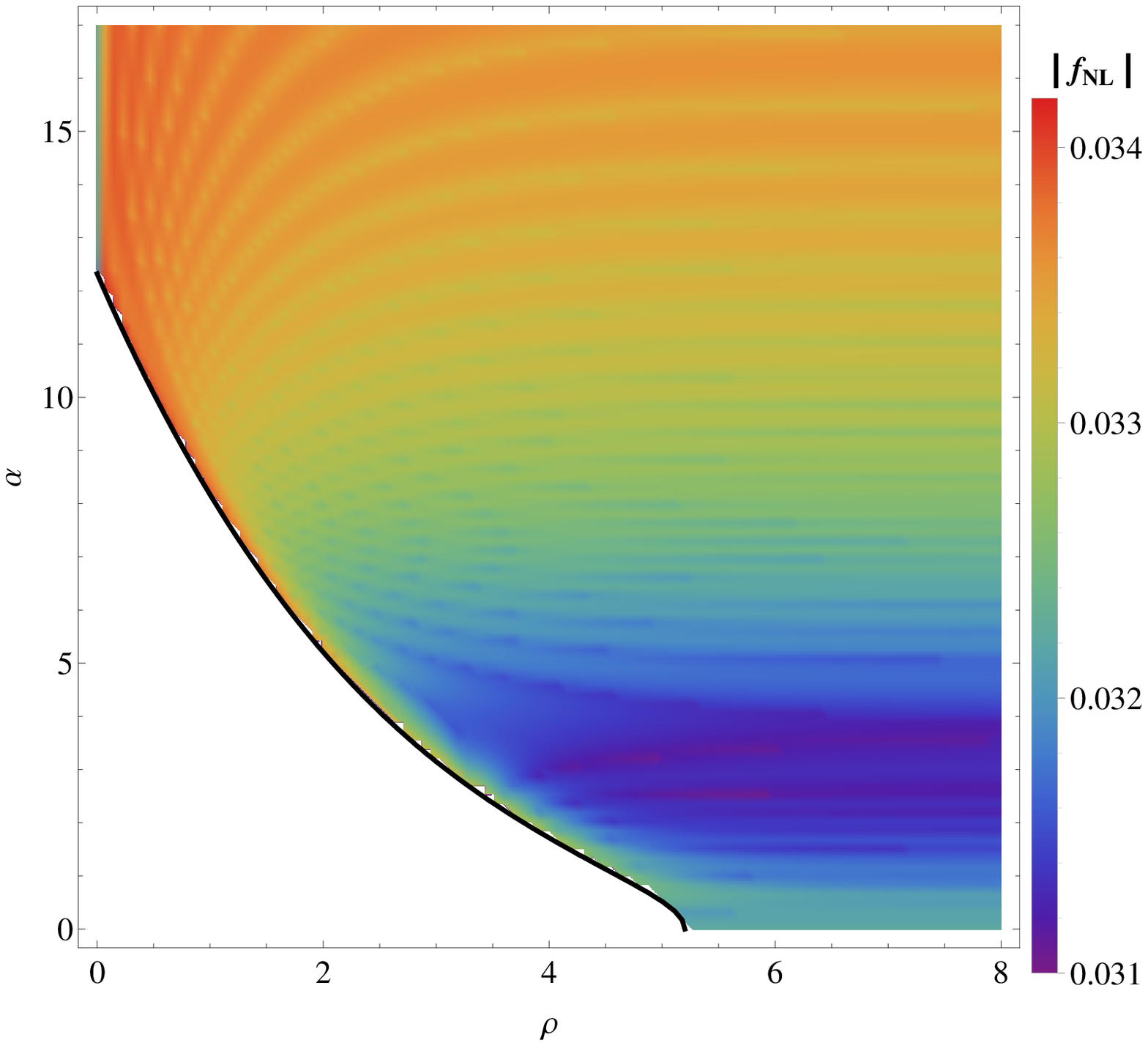}}
	\scalebox{0.49}{\includegraphics{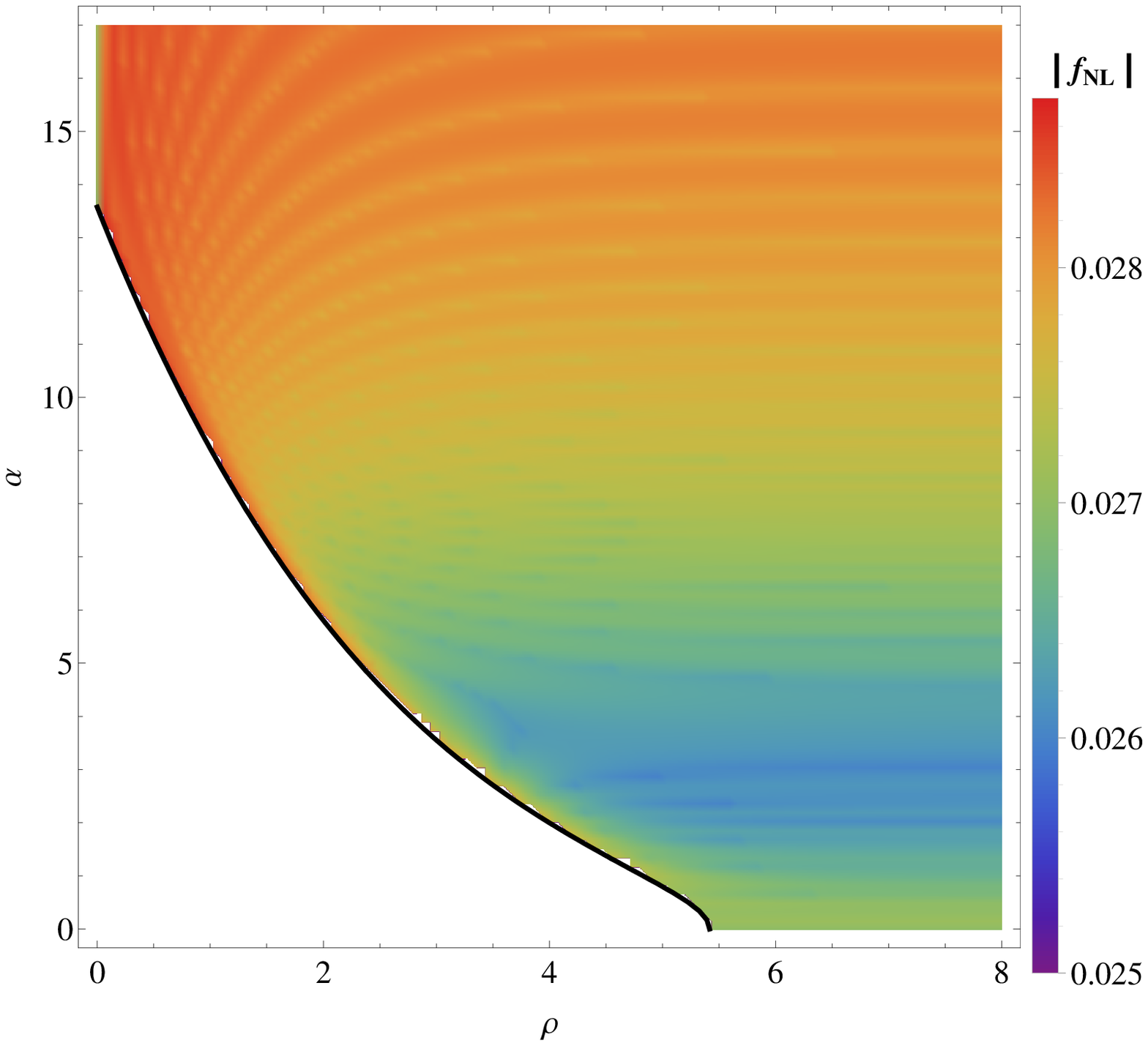}} 
	\caption{\it Values of the non-Gaussianity parameter $f_{\rm NL}$ in the $(\rho,\alpha)$ plane,
	evaluated for $N =50$ (left) and $60$ (right) e-folds, 
	for $c=100$ (upper panel) and $c = 0.1$ (lower panel).} \label{fnlplane}
\end{figure}

The two-field formalism used above relies on the assumption that the scalar field fluctuations are nearly Gaussian,
an assumption that clearly needs to be verified. A quantitative measure of non-Gaussianity is provided by the
dimensionless nonlinear parameter $f_{\rm NL}$, which is related to the bispectrum by~\cite{Komatsu:2001rj}
\beq
\mathcal{B}_{\mathcal{R}}(k_1,k_2,k_3) \; \equiv \; -\frac{6}{5}f_{\rm NL}(k_1,k_2,k_3)[\mathcal{P}_{\mathcal{R}}(k_1)\mathcal{P}_{\mathcal{R}}(k_2) + {\rm cyclic}]\,.
\eeq
For bispectrum configurations of the local type (e.g., $k_1\simeq k_2\gg k_3$), 
$f_{\rm NL}$ can be calculated using the $\delta N$ formalism~\cite{Lyth:2005fi,Peterson:2010np,Peterson:2010mv}.
Namely, for Lagrangian of the form
\beq
\mathcal{L} \; = \; \frac{1}{2}G_{ij}\partial_{\mu}\phi^i\partial^{\mu}\phi^j - V(\phi^1,\phi^2)
\eeq
with the gradients with respect to the fields defined as
\beq
\nabla \; \equiv \; (\partial_i) \ , \ \ \nabla^{T}\equiv (G^{ij}\partial_j) \, ,
\eeq
one finds that 
\beq\label{fnlgen}
f_{\rm NL} \; = \; -\frac{5}{6}\frac{\nabla N \nabla^T\nabla N \nabla^T N}{|\nabla N|^4}\, ,
\eeq
where $N$ is the number of e-foldings, and the gradients are evaluated at horizon exit.
In our particular case, in the non-canonical basis
\beq
T=\frac{1}{\sqrt{2}}(x + i y)
\eeq
the expression (\ref{fnlgen}) conveniently reduces to
\beq\label{fnlpar}
f_{\rm NL} = -\frac{5}{6}\frac{N_{ij}N_i N_j}{(N_i^2)^2} \, .
\eeq
Some results for $f_{\rm NL}$ in the $(\rho, \alpha)$ planes
for $c=100$ and $c = 0.1$ are displayed in Fig.~\ref{fnlplane}~\footnote{We show
here the linear interpolation of the numerical results obtained for a grid of 100$\times$100 points in this plane.
We use the surface defined by by the triplets $\{x,y,N\}$
to calculate the gradients with respect to $N$.}.

The upper panels in Fig.~\ref{fnlplane} show the values of $f_{\rm NL}$ in the $(\rho, \alpha)$ plane
for $c=100$, and the lower panels show results for $c = 0.1$. The left (right) panels assume
a horizon exit 50 (60) e-folds before the end of inflation.
Since, as we saw previously, two-field effects are generally small for $c = 100$,
with no significant enhancement of the scalar power spectrum
and values of $n_s$ and $r$ that are generally close to those in the single-field limit, we do not expect
large values of $f_{\rm NL}$ in the upper panels. Indeed, we do find values of $f_{\rm NL}$ that are very
small, typically $\lesssim 0.015$ in most of the $(\rho,\alpha)$ plane, and attaining values only as large as 0.036
in the limit $\theta \to \pi/2$ corresponding to initial conditions with $\alpha \simeq 0$. This implies that two-field effects are stronger in this limit. This coincides with the behaviour of the power spectrum shown in Fig.~(\ref{thps}) and the shape of the inflationary trajectory near $\theta=\pi/2$ shown in Fig.~(\ref{thtraj}).
Interestingly, as seen in the lower panel of Fig.~\ref{fnlplane}, we also find small values of
$f_{\rm NL} \lesssim 0.032$ even in the case $c = 0.1$ where we might have expected two-field effects
to be much more important. We conclude that, for all points in the $(\rho,\alpha)$ plane in the cases
studied, the assumption of near-Gaussianity is acceptable.

\section{Summary and Conclusions}

We have presented in this paper a general formalism for the study of two-field effects in 
supergravity models. We have specialized to a particular example in the context of
no-scale supergravity motivated by Planck and BICEP2 data. It is known
that such models require terms in the K\"ahler potential that stabilize some
field components, and we have studied how sensitive the predictions for the
CMB observables $n_s, r$ are to their magnitude. We have also studied the
magnitude of the non-Gaussianity parameter $f_{\rm NL}$ in the presence
of two-field effects. Our studies have been within the framework of one
particular form for the stabilization terms, but we expect them to have
broader validity.

We have found that varying the magnitude of the stabilization terms
yields predictions for the tensor-to-scalar ratio $r$ that interpolate
between large BICEP2-friendly values $\sim 0.15$ to much smaller Planck-friendly values.
This is possible along a fixed direction in field space, whereas in the single-field
approximation (which applies when the stabilization is strong) such a BICEP2-Planck
interpolation is possible only by varying the inflationary direction in field space~\cite{EGNO2}.
The new degree of flexibility in the two-field case is due to the enhancement
$\mathcal{P}_{\mathcal{R}}$ of the scalar power spectrum in the two-field case.
This mechanism may offer a way of resuscitating quadratic chaotic inflationary models,
which would otherwise yield large values of $r$ that seem to be disfavoured~\cite{CHW} by the
recent Planck data on dust in the BICEP2 field of view~\cite{Planckdust},
if they are embedded in a broader framework.

In general, our two-field analysis yields values of the scalar spectral tilt
$n_s$ that are highly compatible with the range currently favoured by experiment.
It was known previously that the no-scale supergravity model yields acceptable
values of $n_s$ when treated in the single-field approximation appropriate for
large stabilization terms. It is reassuring that this concordance is not lost
when their magnitude is reduced.

It is also reassuring that our two-field analysis yields value of $f_{\rm NL}$ that
are within the experimental upper bounds. Indeed, our small results for $f_{\rm NL}$
give little encouragement that this type of non-Gaussianity could be measured
within the foreseeable future.

Finally, we note that the analysis made here can easily be extended to other
supergravity models of inflation. As mentioned earlier, a multi-field treatment
is likely to be required for such models, since they are generally described
by effective theories with more than one field. Our results suggest that, on
the one hand, such models are likely capable of yielding results compatible
with current CMB observations at least as long as any stabilization terms have
characteristic energy scales larger than the inflation scale. On the other hand, our results show how
models that appear to yield large values of $r$ in the single-field approximation
can yield much smaller values of $r$ when two-field effects become
important. This feature may become welcome if dust effects are as significant for
the interpretation of the BICEP2 data as has been suggested in~\cite{debate,Planckdust}.
We look forward to the promised upcoming joint publication by the BICEP2 and Planck
collaborations, and to data from other experiments.

\section*{Acknowledgements}

The work of J.E. was supported in part by the London Centre for Terauniverse Studies
(LCTS), using funding from the European Research Council via the Advanced Investigator
Grant 267352 and from the UK STFC via the research grant ST/J002798/1.
The work of D.V.N. was supported in part by the DOE grant DE-FG03-
95-ER-40917, and he would like to thank Andriana Paraskevopoulou for inspiration. 
The work of M.A.G.G. and
K.A.O. was supported in part by DOE grant DE-SC0011842  at the University of
Minnesota.

\end{document}